\documentclass[a4paper,11pt]{article}  
\usepackage{mathptmx}
\usepackage{bm}
\usepackage{berasans}
\usepackage{beramono}

\usepackage[T1]{fontenc}
\usepackage[latin9]{inputenc}
\usepackage{geometry}
\usepackage{color}
\usepackage[english]{babel}
\usepackage{booktabs}
\usepackage{float}
\usepackage{enumerate}
\usepackage{amsthm}
\usepackage{amsmath}
\usepackage{amssymb}
\usepackage{graphicx}
\usepackage{esint}
\usepackage{multirow}
\usepackage{tikz}
\usetikzlibrary{shapes, arrows, positioning, calc}

\usepackage{amsmath,amsfonts}
\usepackage{graphicx}
\usepackage[colorlinks=true, allcolors=blue]{hyperref}
\usepackage[numbers,sort&compress]{natbib}
\usepackage{graphicx} 

\theoremstyle{definition}
\newtheorem{defn}{\protect\definitionname}[section]
\theoremstyle{plain}

\theoremstyle{definition}
\newtheorem{example}{\protect\examplename}[section]
\ifx\proof\undefined\
\newenvironment{proof}[1][\proofname]{\par
	\normalfont\topsep6\p@\@plus6\p@\relax
	\trivlist
	\itemindent\parindent
	\item[\hskip\labelsep
	\scshape
	#1]\ignorespaces
}{%
	\endtrivlist\@endpefalse
}
\providecommand{\proofname}{Proof}
\fi
\theoremstyle{plain}

\theoremstyle{plain}
\newtheorem{rem}{\protect\remarkname}[section]
\theoremstyle{plain}
\newtheorem{thm}{\protect\theoremname}[section]
\theoremstyle{plain}

\newtheorem{property}{Property}[section]
\tikzstyle{startstop} = [ellipse, draw, text width=4em, text centered, minimum height=2em]
\tikzstyle{process} = [rectangle, draw, text width=13em, text centered, minimum height=3em]
\tikzstyle{decision} = [diamond, draw, text width=10em, text centered, minimum size=2em]
\tikzstyle{arrow} = [thick,->,>=stealth,line width=1.2pt]

\usepackage{indentfirst}

\makeatother

\providecommand{\corollaryname}{Corollary}
\providecommand{\definitionname}{Definition}
\providecommand{\examplename}{Example}
\providecommand{\lemmaname}{Lemma}
\providecommand{\propositionname}{Proposition}
\providecommand{\remarkname}{Remark}
\providecommand{\theoremname}{Theorem}

\title{Dynamic stochastic multi-attribute group decision-making based on fractional-order stochastic dominance criteria}

\author{\textit{Shuqin Wang$^{\rm a}$,\ \  Wei Zou$^{\rm b*}$, \ \ Jiehua Xie$^{\rm a}$\thanks{Corresponding
			authors.
			\textit{Email-address:} zouwei@juwp.edu.cn (Wei Zou); jhxie@pku.edu.cn(Jiehua Xie).} }
	\\
	\small  \textit{$^{\rm a}$School of Business Administration, Jiangxi University of Water Resources and Electric Power, Jiangxi 330099, P.R.China}
	\\
	\small  \textit{$^{\rm b}$School of Science, Jiangxi University of Water Resources and Electric Power, Jiangxi 330099, P.R.China}
}

\begin{document}
\maketitle
\begin{abstract}
The stochastic dominance criterion has been widely employed in stochastic multi-attribute decision making as a robust tool for quantitatively comparing decision-makers' preferences over alternatives. Existing research predominantly relies on integer-order stochastic dominance criteria within static modeling frameworks. However, the integer-order scale is often too coarse, and real-world decision-making typically occurs in dynamic environments, for which static methods are inadequate. To address these limitations, this paper proposes a novel dynamic stochastic multi-attribute group decision making model based on fractional-order stochastic dominance criteria. In this model, we introduce the priority degree under fractional-order stochastic dominance, which not only encompasses the priority degree under integer-order stochastic dominance, but also is applicable to decision-makers whose utility functions are not concave everywhere. Consequently, the proposed model captures the dominance relations between alternatives with greater accuracy and flexibility. Moreover, the model allows for the addition or removal of attributes and alternatives across decision periods, thereby exhibiting the applicability to any dynamic decision process. Numerical examples and comparative analyses show the effectiveness and robustness of the proposed approach, highlighting its broader applicability compared with conventional methods. This study enriches the theoretical foundation of dynamic stochastic multi-attribute group decision making and contributes to improving the reliability and precision of decision outcomes.

\end{abstract}
%

\section{Introduction}

Decision problems with multiple alternatives and multiple attributes (multi-attribute group decision-making, MAGDM) arise in supply chain management, engineering project evaluation, and financial investment \cite{wenxian1, wenxian2, wenxian3, wenxian4, wenxian5}. Attribute values have been modeled as interval numbers, linguistic variables, and Z-numbers \cite{wenxian6, wenxian7, wenxian8}. However, in many settings the evaluations of some or all attributes are inherently random and are modeled as stochastic variables, namely stochastic multi-attribute group decision-making (SMAGDM) \cite{wenxian9}. Decisions also unfold over time rather than at a single point, since factors such as environmental changes and policy adjustments alter the problem; this motivates dynamic multi-attribute group decision-making (DMAGDM) \cite{wenxian10, wenxian11, wenxian12}. In such problems, alternatives and attributes may be added or removed in any period rather than fixed in advance. Results from earlier periods also affect the current decision. Static MAGDM methods do not capture these cross-period relationships. Dynamic stochastic multi-attribute group decision-making (DSMAGDM) combines both features: multiple decision-makers (DMs) evaluate alternatives over multiple periods, and the attribute values are stochastic. Such problems arise in management and engineering applications \cite{wenxian13}.

Stochastic dominance (SD) is a standard tool for SMAGDM, it ranks alternatives without requiring the decision makers to spell out explicit utility functions \cite{wenxian14}. Chen et al. \cite{wenxian15} incorporated decision makers' psychological preferences into a hybrid stochastic multi-attribute decision-making method, defining a stochastic dominance degree. The method defines a stochastic dominance degree for different preferences. It draws on the economic interpretation of mean and variance.

Existing stochastic multi-attribute decision-making models use only integer-order stochastic dominance criteria, such as first-order and second-order dominance. M\"uller et al. \cite{wenxian16} point out a limitation of these criteria: their measurement scale is coarse, since every integer order above 1 applies exclusively to risk-averse decision-makers. In practice, however, individuals often act in both risk-averse and risk-seeking ways, buying insurance while investing in stocks, so their risk attitudes are not fully risk-averse. M\"uller et al. \cite{wenxian16} and Huang et al. \cite{wenxian17} studied how dominance can pass continuously from the first to the second order, opening the line of research on fractional-order stochastic dominance. Fractional-order stochastic dominance has since been studied in decision science and risk management \cite{wenxian18, wenxian19}.

Traditional stochastic dominance criteria cannot determine the dominance relation for some pairs of alternatives. The result is either a missing dominance relation or an ambiguous ranking in which some alternatives cannot be ordered, which limits practical use. Leshno and Levy \cite{wenxian20} proposed almost stochastic dominance to address this. Since then, almost stochastic dominance has been used to quantify dominance relations between alternatives and to build SMADM models. Jiang et al. \cite{wenxian21}, for instance, combined it with integer-order stochastic dominance in a method for SMADM. However, methods based on almost stochastic dominance are not suitable for decision-makers with locally concave-convex utility functions, and their quantitative comparison of stochastic attributes remains coarse.

However, most existing studies on multi-attribute decision-making assume a static environment. In practice, such problems unfold over time and require continuous adjustment as conditions change. Campanella and Ribeiro \cite{wenxian22} build a dynamic MADM framework. The framework couples successive periods through a feedback loop that revises how information is associated and aggregated. In this framework, the sets of alternatives and attributes are not fixed: both can evolve and be added or removed across periods. For dynamic SMADM, Hu and Yang \cite{wenxian23} drew on cumulative prospect theory and set pair analysis, illustrating their method with a risk-investment example.

This study develops a DSMAGDM method based on fractional-order stochastic dominance criteria. It makes three contributions. (1) We integrate fractional-order stochastic dominance criteria into a dynamic decision framework that handles stochastic information changing over time. (2) Fractional-order stochastic dominance sharpens the discrimination of integer-order criteria and yields finer distinctions between alternatives. (3) We compare the method with static SMAGDM and with methods using other SD criteria.

The rest of the paper proceeds as follows. Section 2 reviews fractional-order stochastic dominance criteria and the priority degree. Section 3 presents the DSMAGDM method. Section 4 gives a numerical example. Section 5 reports comparative analyses. Section 6 summarizes the findings and points to future work.

\section{Fractional-order stochastic dominance and the priority degree }

Stochastic dominance (SD) rules are one of the most commonly used theoretical tools for identifying the risk preference behavior of decision-makers, as well as a standard tool for the selection and quantitative comparison of risk sets. Although these criteria have been widely  studied and applied, recent research by M\"uller et al. \cite{wenxian16}  identified a shortcoming in the existingSD rules: the integer-order is relatively coarse. For example, first-degree stochastic dominance (FSD) requires the utility function to be increasing, and second-degree stochastic dominance (SSD)  requires it to be increasing and concave. In practice, most decision-makers may be risk averse at certain wealth levels and risk seeking at others, and their utility functions exhibit local convexity. Moving from FSD to SSD is thus a substantial leap, and preference relations between the two orders deserve attention, a continuum of dominance rules rather than a discrete jump. In order to characterize the consensus in ranking distributions for utility functions exhibiting local convexities, M\"uller et al. \cite{wenxian16} interpolated between FSD and SSD. Huang et al. \cite{wenxian17} then introduced a new fractional degree of stochastic dominance for degrees of 1 or larger. Owing to its strong practical background and important theoretical properties, fractional-order stochastic dominance is currently one of the most cutting-edge research directions in fields such as decision science and financial economics. This paper introduces the fractional-order stochastic dominance criteria into the study of DSMAGDM problems, and constructs a priority degree that measures how strongly one alternative dominates another.

\subsection{Fractional-order stochastic dominance criterion}
Assume that the decision-maker's (DM's) utility function \(u(x)\) is continuously differentiable. M\"uller et al. \cite{wenxian16} define the set of applicable utility functions for the DM as
\[
U_{1+\gamma} = \left\{ u(\cdot) \mid 0 \leq \gamma u'(y) \leq u'(x), \, \forall x \leq y \right\}, \quad 0 \leq \gamma \leq 1.
\]

Note that in the above definition, as $\gamma$ approaches 0, the constraint is simplified as $u'(x) \geq 0$, $\forall x$, and thus $U_1$ is the set of all non-decreasing utility functions, which is applicable to all nonsatiable DMs. As $\gamma$ approaches 1, the constraints reduce to $u'(x) \geq 0$ and $u''(x) \leq 0$, $\forall x$, and thus $U_2$ is the set of all increasing and concave utility functions, which is applicable to all nonsatiable and risk-averse DMs. For $0 < \gamma < 1$, $U_{1+\gamma}$ is the set of all increasing utility functions exhibiting local convexities, which is applicable to all nonsatiable DMs who are predominantly risk averse but do not categorically dislike all forms of risk. Moreover, from the definition of $U_{1+\gamma}$, it follows that 
\begin{equation}\label{equ1}
U_2 \subseteq U_{1+\gamma_2} \subseteq U_{1+\gamma_1} \subseteq U_1, \quad 0 \leq \gamma_1 \leq \gamma_2 \leq 1.
\end{equation}

Let $a$ and $b$ be two real numbers such that $a<b$. Suppose that $X$ and $Y$ are two random variables defined on the interval $[a,b]$, and $F$ and $G$ are cumulative distribution functions (cdfs) of $X$ and $Y$, respectively, such that $F(a) = G(a) = 0 \ \text{and} \ F(b) = G(b) = 1$. Let $\mathbb{E}_F[\cdot] \ \text{and} \ \mathbb{E}_G[\cdot]$ be the expected values for distributions $F$ and $G$, respectively. M\"uller et al. \cite{wenxian16} defined a class of fractional-order stochastic dominance criteria, namely the $1+\gamma$-order stochastic dominance criteria, using the utility function set $U_{1+\gamma}$ and the idea of interpolation.

\begin{defn}\label{DEF1}
    (M\"uller et al. \citep{wenxian16}) For $0 \leq \gamma \leq 1$, $F$ dominant $G$ by $1+\gamma$-order stochastic dominance ($1+\gamma$-SD), denoted $F \succeq_{1+\gamma-SD} G$ (or $X \succeq_{1+\gamma-SD}  Y$), if  $\mathbb{E}_F[u(X)] \geq \mathbb{E}_G[u(Y)]$ holds for all $u \in U_{1+\gamma}$.
\end{defn} 
From the definition of $1+\gamma$-SD rule and the set \(U_{1+\gamma}\), it follows that when \(\gamma=0\), \(F \succeq_{1\text{-SD}} G\) corresponds to FSD, i.e., \(F \succeq_{FSD} G\); when \(\gamma=1\), \(F \succeq_{2\text{-SD}} G\) corresponds to SSD, i.e., \(F \succeq_{SSD} G\). The following property presents the relationships among the \(1+\gamma\)-SD, FSD, and SSD criteria.

\begin{property}\label{PRO1}
(M\"uller et al. \citep{wenxian16}) For two cdfs \(F\) and \(G\), the following results hold.
\begin{enumerate}    
    \item[(i)] For any \(0 \leq \gamma_1 \leq \gamma_2 \leq 1\), $F \succeq_{1+\gamma_1-SD} G \implies F \succeq_{1+\gamma_2-SD} G$.
    \item[(ii)] For any \(0 \leq \gamma \leq 1\), $F \succeq_{\text{FSD}} G \implies F \succeq_{1+\gamma\text{-SD}} G \implies F \succeq_{\text{SSD}} G$.
\end{enumerate}
\end{property}

\begin{proof}
\noindent

(i) If $F \succeq_{1+\gamma_1 -SD} G$, then by Definition \ref{DEF1}, we know that $\mathbb{E}_F[u(X)] \geq \mathbb{E}_G[u(Y)]$ holds for all $u \in U_{1+\gamma_1}$. Also, since $U_{1+\gamma_2} \subseteq U_{1+\gamma_1}$ due to $0 \leq \gamma_1 \leq \gamma_2 \leq 1 $, which follows from Equation \eqref{equ1}, then the inequality $\mathbb{E}_F[u(X)] \geq \mathbb{E}_G[u(Y)]$ holds for all $u \in U_{1+\gamma_2}$, and thus $F \succeq_{1+\gamma_2 \text{-SD}} G$.
	
(ii) For any $0 \leq \gamma \leq 1$, it follows from Equation \eqref{equ1} that $U_2 \subseteq U_{1+\gamma_2} \subseteq U_1$. Then, similar to the proof of the statement (i), we obtain the desired results. \end{proof}

A distributional form of the $1+\gamma$-SD criterion is recorded next.

\begin{thm}\label{THM1}
(M\"uller et al. \citep{wenxian16}) For two cdfs $F$ and $G$, $F \succeq_{1+\gamma\text{-SD}} G$ if and only if for all $t \in [a, b]$,
\begin{equation}\label{equ2}
\int_a^t \bigl(F(x)-G(x) \bigr)_+ \, \mathrm{d}x \leq \gamma \int_a^t \bigl(G(x)-F(x) \bigr)_+ \, \mathrm{d}x
\end{equation}
holds, where $(x)_+ = \max\{x,0\}$.
\end{thm}

M\"uller et al. \cite{wenxian16} restricted the ratio of marginal utilities, Huang et al. \cite{wenxian17} instead constrained the degree of absolute risk aversion and obtained a fractional degree of stochastic dominance for degrees of 1 or larger. Their criterion is defined as follows.

Consider a von Neumann-Morgenstern monetary utility function \(u\), which is twice continuously differentiable and satisfies $u'>0$. The Arrow--Pratt measure of absolute risk aversion corresponding to $u$ is defined as
\[
\rho_{u,2}(x) = -\frac{u''(x)}{u'(x)}.
\]

For a constant $c\in(0,1]$, the family of utility functions is
\[
\mathcal{U}_{1,c}= \Big\{u\,\Big|\, u'(x) > 0,\ \text{and } \rho_{u,2}(x) \ge -\left(\frac1c - 1\right),\ \forall x\in[a,b]\Big\}.
\]
When $c$ approaches zero, $\mathcal{U}_{1,0}$ includes all non-satiated utility functions. As $c$ rises and approaches 1, only non-satiated, risk-averse decision-makers belong to $\mathcal{U}_{1,1}$. Therefore, for any $c$ in the open interval $(0,1)$, the underlying preferences may exhibit risk-loving features, yet such risk preference is bounded. With an appropriate selection of $c$, the set $\mathcal{U}_{1,c}$ a variety of utility functions featuring local convex segments. Moreover, from the definition of $\mathcal{U}_{1,c}$, it follows that $\mathcal{U}_{1,1} \subset \mathcal{U}_{1,c} \subset \mathcal{U}_{1,0}
$

\begin{defn}\label{DEF2}
	(Huang et al. \citep{wenxian17}) For $0 < c \le 1$, $F$ dominates $G$ by $1+c$-order stochastic dominance ($1+c$-SD), denoted $F \succeq_{1+c\text{-SD}} G$ (or $X \succeq_{1+c\text{-SD}} Y$), if $\mathbb{E}_F\left[u(X)\right] \ge \mathbb{E}_G\left[u(Y)\right]$ holds for all $u \in \mathcal{U}_{1,c}$.
\end{defn}

The $1+c$-SD criterion also admits an equivalent condition in terms of the distributions.
\begin{thm}\label{THM2}
	(Huang et al. \citep{wenxian17}) Denote the cdfs of $X$ and $Y$ by $F$ and $G$, respectively. For $c\in(0,1]$, $X$ $(1+c)$SD $Y$ if and only if
	\begin{equation}\label{equ3}
		\int_{a}^{x}\big[G(t)-F(t)\big]dh_c(t)\ge 0,\quad \text{for all } x\in[a,b]
	\end{equation}
	holds, where
\[
h_c(x)=
\begin{cases}
	e^{(1/c-1)x}, & \text{if } 0<c<1,\\
	x, & \text{if } c=1.
\end{cases}
\]
\end{thm}
From the theorem of $1+c$-SD rule and the set $\mathcal{U}_{1,c}$, it follows that when \(c \to 0\), \(F \succeq_{1\text{-SD}} G\) corresponds to FSD, i.e., \(F \succeq_{FSD} G\); when \(c=1\), \(F \succeq_{2\text{-SD}} G\) corresponds to SSD, i.e., \(F \succeq_{SSD} G\). The following property presents the relationships among the \(1+c\)-SD, FSD, and SSD criteria.

\begin{property}\label{PRO2}
	(Huang et al. \citep{wenxian17}) For two cdfs \(F\) and \(G\), the following results hold.
	\begin{enumerate}    
		\item[(i)] For any \(0 < c_1 \leq c_2 \leq 1\), $F \succeq_{1+c_1-SD} G \implies F \succeq_{1+c_2-SD} G$.
		\item[(ii)] For any \(0 < c \leq 1\), $F \succeq_{\text{FSD}} G \implies F \succeq_{1+c\text{-SD}} G \implies F \succeq_{\text{SSD}} G$.
	\end{enumerate}
\end{property}

In contrast to the approach of Huang et al. \cite{wenxian17}, which add constraints on the degree of absolute risk aversion, Mao et al. \cite{wenxian19} proposed a new fractional degree stochastic dominance, which add constraints on the degree of relative risk aversion.

Consider a von Neumann-Morgenstern monetary utility function \(u\), which is twice continuously differentiable and satisfies $u'>0$. The Arrow--Pratt measure of relative risk aversion corresponding to $u$ is defined as
\[
R_{u,2}(x) = -\frac{xu''(x)}{u'(x)}.
\]

For a constant $\ell \in(0,1]$, the family of utility functions is
\[
\mathcal{U}_{1,\ell}= \Big\{u\,\Big|\, u'(x) > 0,\ \text{and } R_{u,2}(x) \ge -\left(\frac1c - 1\right),\ \forall x\in(0,\infty)\Big\}.
\]

\begin{defn}\label{DEF3}
	(Mao et al. \cite{wenxian19}) For $0 < \ell \le 1$, $F$ dominates $G$ by $1+c$-order stochastic dominance ($1+\ell$-SD), denoted $F \succeq_{1+\ell \text{-SD}} G$ (or $X \succeq_{1+\ell \text{-SD}} Y$), if $\mathbb{E}_F\left[u(X)\right] \ge \mathbb{E}_G\left[u(Y)\right]$ holds for all $u \in \mathcal{U}_{1,\ell}$.
\end{defn}

Mao et al. \cite{wenxian19} likewise characterized $1+\ell$-SD through the distribution functions. 
\begin{thm}\label{THM3}
	(Mao et al. \cite{wenxian19}) Denote the cdfs of $X$ and $Y$ by $F$ and $G$, respectively. For $\ell \in(0,1]$, $X$ $(1+\ell)$SD $Y$ if and only if
	\begin{equation}\label{equ4}
		\int_{0}^{x}\big[G(t)-F(t)\big]dh_c(t)\ge 0,\quad \text{for all } x\in(0,\infty)
	\end{equation}
	holds, where $h_\ell (x)=x^{1/ \ell}$, $0 < \ell \le 1$.
\end{thm}

Any property satisfied under the $1+c$ stochastic dominance is valid for the $1+\ell$ stochastic dominance as well. The corresponding discussion is omitted for brevity.

\subsection{The priority degree under fractional-order stochastic dominance}
The priority degree is a theoretical tool used to measure the degree of dominance between two alternatives (random variables) \cite{wenxian24}. In the literature, many priority degrees under different SD rules have been introduced and  applied to SMADM \cite{wenxian25}. The existing priority  degrees are all integer-order. However, the scale of integer-order is too coarse. The jump from priority degree under FSD to priority degree under SSD is
substantial. For any order greater than one, they are only applicable for risk-averse DMs when measuring the preference degree of alternatives. To overcome this limitation, we establish a continuum of priority degrees under SD.

\begin{defn}\label{DEF4}
For $0 \leq d \leq 1$, the priority degree that $F$ is superior to $G$ under $1+\gamma$-SD($1+c$-SD, $1+\ell
$-SD) rule is defined as
\begin{equation}\label{equ5}
\begin{split}
	\mathbb{D}\big[F \succeq_{1+d\text{-SD}} G\big]
	&= \frac{(1-d)\cdot\left[\int_{a}^{b} (1-F(x))\mathrm{d}x - \int_{a}^{b} (1-G(x))\mathrm{d}x\right]}{\int_{a}^{b} (1-F(x))\mathrm{d}x} \\
	&\quad + \frac{d\cdot\left[\int_{a}^{b} (1-F(x))^2\mathrm{d}x - \int_{a}^{b} (1-G(x))^2\mathrm{d}x\right]}{\int_{a}^{b} (1-F(x))^2\mathrm{d}x},\quad 0\le d\le 1
\end{split}
\end{equation}
\end{defn}
If the $1+\gamma$-order stochastic dominance relation exists, $d=\gamma$; if the $1+c$-order stochastic dominance relation exists, $d=c$; if the $1+\ell$-order stochastic dominance relation exists, $d=\ell$.

By comparative analysis, the distinctions and relationships between the priority degree under $1+\gamma$ ($1+c$, $1+\ell$)-order fractional-order stochastic dominance and the existing priority degrees under stochastic dominance can be obtained.

\begin{rem}\label{REM1}
For random variables $X$ and $Y$, (i) when $d=0$, i.e., $F \succeq_\text{FSD} G$, Equation (5) can be simplified as $\mathbb{D}\big[F \succeq_{1\text{-SD}} G\big]={\left(\int_{a}^{b} (1-F(x))\mathrm{d}x - \int_{a}^{b} (1-G(x))\mathrm{d}x\right)}/{\left(\int_{a}^{b} (1-F(x))\mathrm{d}x\right)}$, since $\mathbb{E}_F[X]=\int_{a}^b (1-F(x)) \mathrm{d}x$ and $\mathbb{E}_G[Y]=\int_{a}^b (1-G(x)) \mathrm{d}x$, so when $d=0$, $\mathbb{D}\bigl[F \succeq_{1+d\text{-SD}} G\bigr]=(\mathbb{E}_F[X]-\mathbb{E}_G[Y])/\mathbb{E}_F[X]$, which is consistent with the priority degree under FSD introduced by Chen et al. \cite{wenxian15} (ii) When $d=1$, i.e., $F \succeq_\text{SSD} G$, Equation (5) can be simplified as $\mathbb{D}\big[F \succeq_{2\text{-SD}} G\big]={\left[\int_{a}^{b} (1-F(x))^2\mathrm{d}x - \int_{a}^{b} (1-G(x))^2\mathrm{d}x\right]}/{\int_{a}^{b} (1-F(x))^2\mathrm{d}x}$. Note that $\int_a^b(1-F(x))^2\mathrm{d}x=\mathbb{E}[\min\{X_1,X_2\}]$, where $X_1$ and $X_2$ are independent random variables distributed identically with $X$. Hence, when $d=1$, the priority degree additionally captures the dispersion of the distributions, and thus measures the strength of the second-order stochastic dominance. (iii) When $0<d<1$, the priority degree is a convex combination of the first-order and second-order degrees, that is, $\mathbb{D}=(1-d)T_1+dT_2$, where $T_1=(\mathbb{E}_F[X]-\mathbb{E}_G[Y])/\mathbb{E}_F[X]$ and  $T_2={\left[\int_{a}^{b} (1-F(x))^2\mathrm{d}x - \int_{a}^{b} (1-G(x))^2\mathrm{d}x\right]}/{\int_{a}^{b} (1-F(x))^2\mathrm{d}x}$. Consequently, the value of $\mathbb{D}$ always lies between $T_1$ and $T_2$. The value of $d$, which is determined by the order of the dominance relation itself, specifies the relative contribution of the first-order and second-order information: the closer $d$ is to $0$, the more the degree relies on the first-order information about the expected values, and the closer $d$ is to $1$, the more it relies on the second-order information about the dispersion. In this way, the priority degree adapts itself to the order of each dominance relation rather than measuring all relations on a uniform scale.
\end{rem}

Using the distributional equivalent condition of $1+\gamma$-SD ($1+c$-SD, $1+\ell
$-SD) in Property \ref{PRO2}, we can obtain the upper and lower bounds of the priority degree under $1+\gamma$-SD ($1+c$-SD, $1+\ell
$-SD).

\begin{property} For two random variables $X$ and $Y$ with cdfs $F$ and $G$ respectively, it holds that 
$0 \leq \mathbb{D}\bigl[F \succeq_{1+d\text{-SD}} G\bigr] \leq 1$.
\end{property}

\begin{proof}

\noindent

By Equation (5), the priority degree can be rewritten as the convex combination $\mathbb{D}=(1-d)\,T_1+d\,T_2$, where
\begin{equation*}
	T_1=\frac{\int_a^b(1-F(x))\mathrm{d}x-\int_a^b(1-G(x))\mathrm{d}x}{\int_a^b(1-F(x))\mathrm{d}x},\qquad
	T_2=\frac{\int_a^b(1-F(x))^2\mathrm{d}x-\int_a^b(1-G(x))^2\mathrm{d}x}{\int_a^b(1-F(x))^2\mathrm{d}x}.
\end{equation*}
Since the priority degree is defined only for pairs of alternatives with an existing dominance relation, we have $F\succeq_{1+d\cdot SD}G$ with $0\le d\le 1$, and it suffices to show that both $0\le T_1\le 1$ and $0\le T_2\le 1$.

First, since $0\le 1-F(x)\le 1$ on $[a,b]$ and $X$ is not degenerate at $b$, both denominators are positive, so $T_1$ and $T_2$ are well defined. Furthermore, since $\int_a^b(1-G(x))\mathrm{d}x\ge0$ and $\int_a^b(1-G(x))^2\mathrm{d}x\ge0$, we immediately obtain $T_1\le1$ and $T_2\le1$.

It remains to prove $T_1\ge0$ and $T_2\ge0$. Since $d\le1$, the relation $F\succeq_{1+d\cdot SD}G \implies F\succeq_{2\cdot SD}G$, because the utility class of the $(1+d)$-order stochastic dominance contains all increasing and concave utility functions \cite{Mullerd2017Between}. By the integral characterization of SSD, it holds that
\begin{equation*}
	S(x):=\int_a^x\bigl(G(t)-F(t)\bigr)\mathrm{d}t\ \ge\ 0,\qquad x\in[a,b].
\end{equation*}
For the first-order term, we have
\begin{equation*}
	\int_a^b(1-F(x))\mathrm{d}x-\int_a^b(1-G(x))\mathrm{d}x
	=\int_a^b\bigl(G(x)-F(x)\bigr)\mathrm{d}x=S(b)\ge0,
\end{equation*}
and hence $T_1\ge0$. For the second-order term, note that
\begin{align*}
	\int_a^b(1-F(x))^2\mathrm{d}x-\int_a^b(1-G(x))^2\mathrm{d}x
	&=\int_a^b\bigl(G(x)-F(x)\bigr)\bigl(2-F(x)-G(x)\bigr)\mathrm{d}x.
\end{align*}
Let $w(x)=2-F(x)-G(x)$. Since $F$ and $G$ are nondecreasing, $w$ is nonincreasing on $[a,b]$ with $0\le w(x)\le2$ and $w(b)=2-F(b)-G(b)=0$. Integrating by parts and using $S(a)=0$, we obtain
\begin{equation*}
	\int_a^b\bigl(G(x)-F(x)\bigr)w(x)\mathrm{d}x
	=S(b)\,w(b)-\int_a^b S(x)\,\mathrm{d}w(x)
	=-\int_a^b S(x)\,\mathrm{d}w(x)\ \ge\ 0,
\end{equation*}
where the integral with respect to $\mathrm{d}w$ is understood in the Riemann--Stieltjes sense, since $S(x)\ge0$ and $\mathrm{d}w\le0$ on $[a,b]$. Hence $T_2\ge0$.

Combining the above results yields $0\le T_1\le1$ and $0\le T_2\le1$; since $d\in[0,1]$, the priority degree, being a convex combination of $T_1$ and $T_2$, satisfies $0\le\mathbb{D}[F\succeq_{1+d\cdot SD}G]\le1$.
\end{proof}

The following example is given to verify the reasonableness of the proposed dominance degree.

\begin{example}
	Consider three alternatives $X$, $Y$ and $Z$, whose probability distributions over the evaluation grades $1,2,\ldots,5$ are listed in Table~\ref{Table:NE1}.
	\begin{table}[htbp]\footnotesize
		\centering
		\caption{Probability distribution information of the alternatives in Example~\ref{Table:NE1}}
		\label{Table:NE1}
		\setlength{\tabcolsep}{6pt}
		\begin{tabular}{lccccc}
			\toprule
			\multirow{2}{*}{Alternative} & \multicolumn{5}{c}{Probability distribution} \\
			\cmidrule(lr){2-6}
			& $1$ & $2$ & $3$ & $4$ & $5$ \\
			\midrule
			$X$ & $0.100$ & $0.120$ & $0.400$ & $0.310$ & $0.070$ \\
			$Y$ & $0.320$ & $0.070$ & $0.390$ & $0.180$ & $0.040$ \\
			$Z$ & $0.240$ & $0.170$ & $0.090$ & $0.220$ & $0.280$ \\
			\bottomrule
		\end{tabular}
	\end{table}
	
	Let $F_1$, $G$ and $F_2$ denote the cdfs of $X$, $Y$ and $Z$, respectively. By the distributional equivalence condition of the fractional-order stochastic dominance given in Equation \eqref{equ2}, the dominance relations among the three alternatives are identified as $F_1\succeq_{FSD}G$ and $F_2\succeq_{1.25\cdot SD}G$, whereas $F_2\succeq_{FSD}G$ does not hold. 
	
	For the pair $(F_1,G)$, by Equation \eqref{equ5}, the priority degrees are computed as $\mathbb{D}[F_1\succeq_{FSD}G]=0.1853$, $\mathbb{D}[F_1\succeq_{1.25\cdot SD}G]=0.2055$. For the pair $(F_2,G)$,  the priority degree computed is $\mathbb{D}[F_2\succeq_{1.25\cdot SD}G]=0.1800.$
	
	(i) Notably, at the same order $1.25$, the priority degree of the first-order dominance relation exceeds that of the fractional-order one, i.e., $\mathbb{D}[F_1\succeq_{1.25\cdot SD}G]=0.2055>\mathbb{D}[F_2\succeq_{1.25\cdot SD}G]=0.1800$. This is consistent with the intuition that $F_1\succeq_{FSD}G$ is a stronger relation than $F_2\succeq_{1.25\cdot SD}G$, and it shows that the proposed degree distinguishes dominance relations that are identical from the first-order viewpoint.
	
	(ii) By Property 1, the relation $F_1\succeq G$ holds at every order, i.e., $F_1\succeq_{FSD}G\Rightarrow F_1\succeq_{1.25\cdot SD}G\Rightarrow F_1\succeq_{1.5\cdot SD}G\Rightarrow F_1\succeq_{SSD}G$, so the priority degree increases as the order rises:
	\begin{equation*}
		\mathbb{D}[F_1\succeq_{1.25\cdot SD}G]=0.2055\
		<\ \mathbb{D}[F_1\succeq_{1.5\cdot SD}G]=0.2257\
		<\ \mathbb{D}[F_1\succeq_{SSD}G]=0.2661.
	\end{equation*}
	
	The above results agree with the intuition that a stronger dominance relation yields a larger degree at the same order, and that the degree adapts itself to the order of the relation; hence the reasonableness of the proposed dominance degree is verified.
\end{example}

\begin{rem}\label{REM2}
The value of priority degree $\mathbb{D}\bigl[F \succeq_{1\text{-SD}} G\bigr]$ under the $1+d$-SD characterizes the strength of $F$ dominating $G$ under the $1+d$-SD criterion. A larger value of $\mathbb{D}\bigl[F \succeq_{1\text{-SD}} G\bigr]$ indicates a stronger degree of dominance.
\end{rem}

\section{Dynamic stochastic MAGDM approach based on $1+\gamma$-SD}

In reality, data information of decision-making may come from different stages or periods, and then decision-making is usually a dynamic process in which early decision results significantly affect subsequent outcomes. Also, due to the uncertainty in which the natural states
faced by DMs are random, attribute values may be expressed as stochastic variables in some practical MAGDM problems. Hence,  the dynamic stochastic MAGDM (thereafter DSMAGDM) problem is a topic of both theoretical and practical interest \cite{wenxian26}.

This section sets out the DSMAGDM model built on fractional-order stochastic dominance. Table \ref{Table:ST1} lists the symbols used throughout the paper.

\begin{table}[htbp]\footnotesize
  \centering
  \caption{Symbols and explanations}
  \label{Table:ST1}
  \setlength\tabcolsep{4pt}
  \begin{tabular}{ll}
    \toprule
    Notations & Explanations \\
    \midrule
$P = \{1,2,3,\cdots\}$ & Index set of the decision periods\\
$A^p = \{A_1^p,A_2^p,\cdots,A_m^p\}$ & Set of m alternatives, $A_i^p$ refers to alternative $i$ at period $p$\\
$C^p = \{C_1^p,C_2^p,\cdots,C_n^p\}$ & Set of n attributes, $C_j^p$ refers to attribute $j$ at period $p$\\
$E^p = \{e_1^p,e_2^p,\cdots,e_r^p\}$ & Set of r experts, $e_f^p$ is expert $f$ at period $p$\\
$w^p = (w_1^p,w_2^p,\cdots,w_n^p)^T$ & Attribute-weight vector, with $w_j^p$ the weight on attribute $j$, satisfying $\sum_{j=1}^n w_j = 1$ and $0\leq w_j \leq 1$
\\
$\boldsymbol{\lambda}^p=(\lambda_1^p,\lambda_2^p,\dots,\lambda_r^p))^T$ & Expert-weight vector, with $\lambda_f^p$ the weight assigned to expert $f$, satisfying $\sum_{f=1}^r \lambda_f = 1$ and $0\leq \lambda_f \leq 1$ \\
$X = \left[ X_{fij}^p \right]_{m \times n}$ & The decision matrix \\
$F_{fij}^p$ & The distribution function of $X_{fij}^p$ \\
$\mu_{fij}^p$ & The mathematical expectation of $X_{fij}^p$ \\
$X^*=\left[ X_{fij}^{p*}\right]_{m \times n}$ & The normalized decision matrix \\
$\widetilde{R}_{fj}^p=\left[\widetilde{r}_{fil}^{pj} \right]_{m \times m}$ & Matrix of the dominance relations \\
$\Phi^p=\left[\varphi_{il}^p \right]_{m \times m}$ & The priority degree matrix \\
$S^{p+}(A_i)$ & Positive flow of $A_i$ at period $p$ \\
$S^{p-}(A_i)$ & Negative flow of $A_i$ at period $p$ \\
$S^p(A_i)$ & Net flow of $A_i$ at period $p$ \\
$\widetilde{S}^p(A_i)$ & Normalized value of the net flow of $A_i$ at period $p$ \\
$\smash{\widetilde{\widetilde{S}}}^{\,p}(A_i)$ & Dynamic normalized value of the net flow of $A_i$ at period $p$ \\
$\mathcal{H}^p$ & Historical set at period $p$ \\
 \bottomrule
  \end{tabular}
\end{table}

\subsection{Description of a DSMAGDM problem }

Consider a DSMAGDM problem that unfolds over periods $p \in P = \{1,2,3,\ldots\}$. At period $p$ the candidate set is $A^p = \{A_1^p, A_2^p, \ldots, A_m^p\}$, with alternatives indexed by $i = 1,\ldots,m$, the attributes are $C^p = \{C_1^p, \ldots, C_n^p\}$, $j = 1,\ldots,n$, and the experts are $E^p = \{e_1^p, \ldots, e_r^p\}$, $f = 1,\ldots,r$. Attribute weights at period $p$ are collected in $w^p = (w_1^p, \ldots, w_n^p)^T$, with $w_j^p \in [0,1]$ and $\sum_{j=1}^n w_j^p = 1$, expert weights are collected in $\lambda^p = (\lambda_1^p, \ldots, \lambda_r^p)^T$, where $\lambda_f^p \in [0,1]$ and $\sum_{f=1}^r \lambda_f^p = 1$. For each triple $(i,j,f)$ at period $p$, expert $e_f^p$ evaluates alternative $A_i$ on attribute $C_j$ by a random variable $\tilde{X}_{fij}^p$, whose distribution function and mean are $F_{fij}^p$ and $\mu_{fij}^p$. These evaluations form the stochastic decision matrix $X = [X_{fij}^p]_{m \times n}$. 

Two questions are addressed. One is to rank the alternatives from $X$; the other is to identify the most preferred alternative by combining earlier-period results with the historical set $\mathcal{H}^p$. The set $\mathcal{H}^p$ follows a top-$K$ rule: the $K$ highest-ranked alternatives of the current period are carried over. An alternative dropped from the current candidate set but still recorded in $\mathcal{H}^p$ keeps its earlier evaluation, so that its history is available again should it re-enter in a later period. $\mathcal{H}^p$ is then updated as
\[\mathcal{H}^p = \mathcal{T}_K\big(O^p\big) \cup \big(\mathcal{H}^{p-1} \setminus A^p\big)\]
where $\mathcal{T}_K(O^p)$ denotes the set of the top-$K$ alternatives in the ranking $O^p$ of the current period. That is, besides the top-$K$ alternatives of the current period, the alternatives that belong to the historical set but leave the alternatives set are retained in the historical set together with their preserved evaluations.

To address the above problem, the DSMAGDM approach based on fractional-order stochastic dominance criteria is proposed in this paper. In the proposed approach, the procedure is divided into two steps.

$(I)$. At any period $p$, we first provide a procedure to identify the dominance relation between any pair of alternatives on the basis of fractional-order stochastic dominance criteria. Then, through the identified fractional-order dominance relations, the priority degrees between any pair of alternatives are obtained. The overall priority degree for each pair of alternatives is further determined according to the weights of attributes and experts. Finally, all alternatives at the period $p$ are ranked according to the net flow value derived from PROMETHEE~II, and the historical set  $\mathcal{H}^p$ in the current period is determined accordingly.

$(II)$. The dynamic nature of the decision process is dealt with by means
of a feedback mechanism. In every period $p$, if an alternative in the current set also belongs to the historical set $\mathcal{H}^{p-1}$  of the previous period, the net flow in this period and that in the previous period should be aggregated into a new net flow according to the aggregation rule, then alternatives are ranked based on the net flow values. The net flow of subsequent periods are calculated in the same manner until the entire dynamic decision process is completed. At each decision period, the historical set is updated in accordance with the retention rule adopted in this study, which selects the $K$ alternatives with the highest rankings to be carried over to subsequent periods. The entire dynamic decision-making procedure continues until the stopping criterion is satisfied, which is determined by the DM based on the specific decision context.

\subsection{The DSMAGDM framework}

The computation proceeds in the steps below.

\noindent\textbf{Step 1. Normalization of stochastic decision matrices.}

In multi-attribute decision-making problems, both benefit-type attributes, whose larger values are better, and cost-type attributes, where the opposite holds, usually coexist. To eliminate the potential impact of different attribute types on subsequent calculation results, normalization of attribute values is required, and the detailed procedure is presented as follows.

If $C_j^p$ is benefit-type attribute, the normalization formula for the attribute evaluation value $X_{fij}^p$ is:
\begin{equation}\label{equ6}
X_{fij}^{p*}=\frac{X_{fij}^p-X_{fij}^{p\min}}{X_{fij}^{p\max}-X_{fij}^{p\min}},\quad(i=1,2,\dots,m;j=1,2,\dots,n;p=1,2,3,\dots).
\end{equation}

If $C_j^p$ is cost-type attribute, the normalization formula for the attribute evaluation value $X_{fij}^p$ is:
\begin{equation}\label{equ7}
X_{fij}^{p*}=\frac{X_{fij}^{p\max}-X_{fij}^p}{X_{fij}^{p\max}-X_{fij}^{p\min}},\quad (i=1,2,\dots,m;j=1,2,\dots,n;p=1,2,3,\dots).
\end{equation}

\noindent Here $X_{fij}^{p\max}$ and $X_{fij}^{p\min}$ are the largest and smallest attribute evaluations at period p. The normalized evaluation value $X_{fij}^{p*}$ remains a random variable, and the processed stochastic decision matrix is $X^*=\left[ X_{fij}^{p*}\right]_{m \times n}$.

\noindent\textbf{Step 2. Determination of stochastic dominance relations.}

\begin{figure}[htbp]
\centering
\begin{tikzpicture}[node distance=1cm, auto]

\node (start) [startstop] {Start};
\node (input) [process, below=of start] {Input $F, G, h$};
\node (k0) [process, below=of input] {$k = 0$};
\node (cal) [process, below=of k0] {$\alpha = 1 + k/h$};
\node (judge1) [decision, below=of cal] {$F \succ_{1+k/h\text{-SD}} G$?};
\node (judge2) [decision, right=of judge1, xshift=2.5cm] {$k \le h$?};
\node (kadd) [process, above=of judge2, yshift=1.1cm] {$k = k + 1$};
\node (yes) [process, below=of judge1, yshift=-0.3cm] {
    $k=0$: FSD \\
    $k=h$: SSD \\
    Others: 1+k/h-SD
};
\node (no) [process, below=of judge2, yshift=-0.3cm] {No SD between FSD and SSD};
\node (end) [startstop, below=of yes, xshift=3.2cm] {End};

\draw [arrow] (start) -- (input);
\draw [arrow] (input) -- (k0);
\draw [arrow] (k0) -- (cal);
\draw [arrow] (cal) -- (judge1);
\draw [arrow] (judge1) -- node[anchor=east] {Yes} (yes);
\draw [arrow] (judge1.east) -- node[anchor=south] {No} (judge2.west);
\draw [arrow] (judge2) -- node[anchor=east] {Yes} (kadd);
\draw [arrow] (kadd.west) -- ($(k0.south)!0.5!(cal.north)$); 
\draw [arrow] (judge2) -- node[anchor=east] {No} (no);
\draw [-] (yes.south) -- ++(0,-1.4);
\draw [arrow, shorten >=3pt] (yes.south |- end.west) -- (end.west);
\draw [-] (no.south) -- ++(0,-2.2);
\draw [arrow, shorten >=3pt] (no.south |- end.east) -- (end.east);

\end{tikzpicture}
\caption{Procedure for determining $1+k/h$-order fractional stochastic dominance}
\label{FIG}
\end{figure}
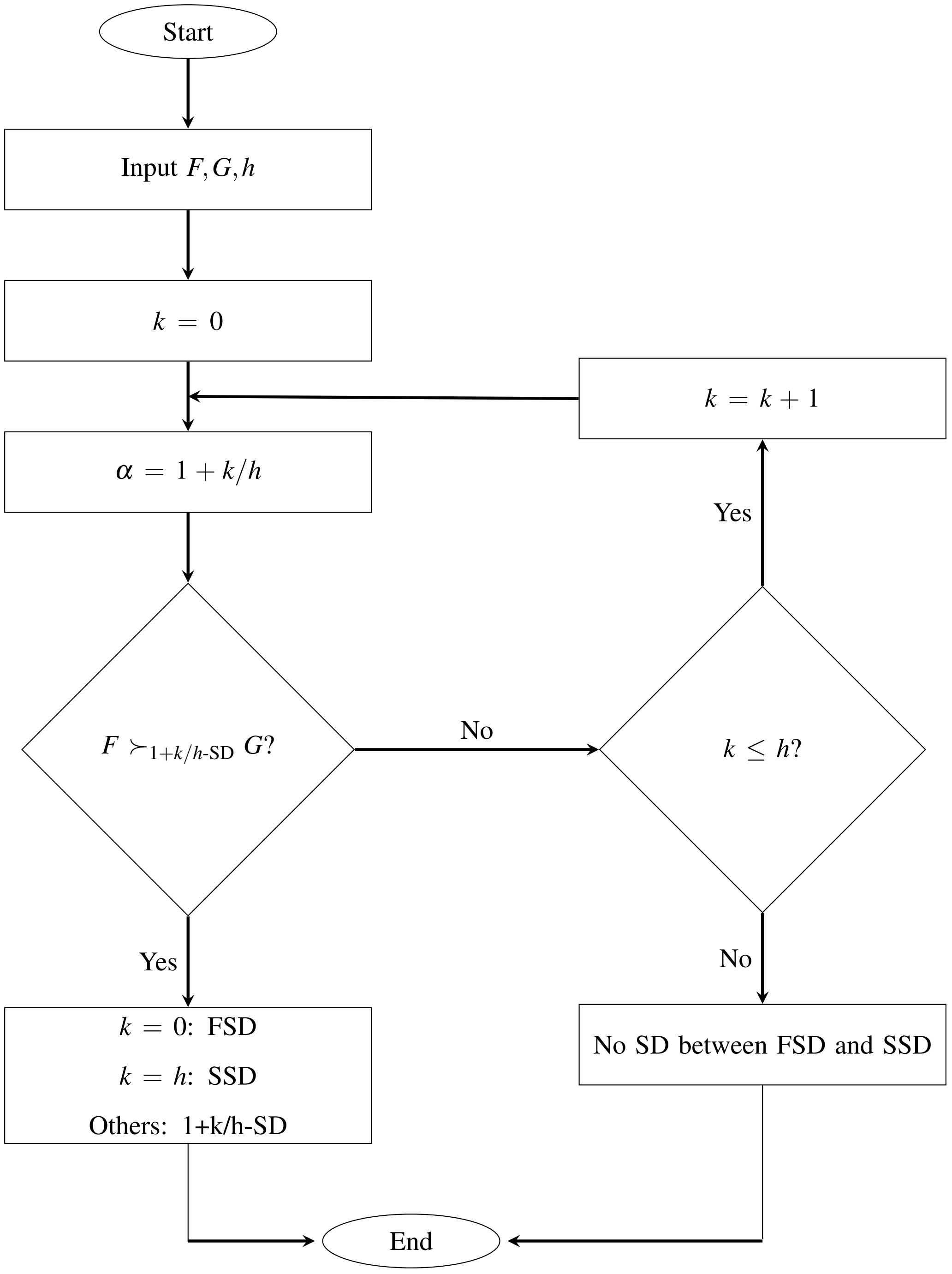

At period $p$, for the alternatives $A_i^p$ , the corresponding distribution function $F_{fij}^p$ is first calculated based on the evaluation values $X_{fij}$ provided by expert $e_f$ for the attribute $C_j^p$. Then, for the same attribute $C_j$, the dominance relation between two alternatives $A_i$ and $A_l$ is determined by Equation \eqref{equ2}. In this study, the $1+\gamma$-order stochastic dominance criterion is adopted to quantify and compare the dominance relations among different alternatives, where the fractional order $\gamma$ can be any constant in $[0,1]$. Fig \ref{FIG} illustrates how the fractional-order dominance relation is identified for every pair of alternatives. Using the procedure, a dominance relation matrix $\widetilde{R}_{fj}^p=\left[\widetilde{r}_{fil}^{pj}\right]_{m \times m}$ can be constructed for each attribute individually.
In this matrix, the element $\widetilde{r}_{fil}^{pj}$ denotes the $1+\gamma$-order stochastic dominance relation between alternatives $A_i$ and $A_l$ under attribute $C_j$ at period $p$, which is defined as:

\[
\widetilde{r}_{fil}^{pj} = 
\begin{cases} 
1+\gamma\text{-SD}, & \text{if } F_{fij}^p \succ_{1+\gamma\text{-SD}} G_{flj}^p, \\
\underline{\quad\quad}, & \text{otherwise}.
\end{cases}
\]

Here, $\gamma =1+k/h$ $(k=0,\dots,h)$ denotes the fractional order, which is obtained by equally dividing the interval [0,1] into $h$ parts.

\noindent\textbf{Step 3. Calculation of priority degrees.}

For alternatives with a fractional-order stochastic dominance relation, the priority degree is calculated to quantify the strength of the dominance relation. Under the $j$th attribute, the priority degree between alternatives $A_i$ and $A_l$ is computed by Equation \eqref{equ3}. To characterize the dominance strength between alternatives more intuitively, the priority degree matrix $\Phi_{fj}^p=\left[\varphi_{fil}^{pj}\right]_{m \times m}$ for the $j$th attribute given by expert $e_f$ in the $p$th period is constructed, where $\varphi_{fil}^{pj}$ denotes the strength of the stochastic dominance relation between alternatives $A_i$ and $A_l$, as defined below:

\[
\varphi_{fil}^{pj} = 
\begin{cases} 
D(F_{fij} \succ_{1+\gamma\text{-SD}} G_{flj}), & \text{if } \widetilde{r}_{fil}^{pj} = 1+\gamma\text{-SD}, \\
0, & \text{otherwise},
\end{cases}
\]

\noindent\textbf{Step 4. Calculation of comprehensive priority degrees.}

To determine the strength of the dominance relationships among all alternatives, it is necessary to calculate the comprehensive priority degree to quantify the overall strength of these dominance relationships. Based on the attribute weight vector $w^p=(w_1^p,w_2^p,\dots,w_n^p)$ and the expert weight vector $\boldsymbol{\lambda}^p=(\lambda_1^p,\lambda_2^p,\dots,\lambda_r^p)$ in the $p$th period, the comprehensive priority degree between alternatives $A_i$ and $A_l$ is computed by weighted summation of the individual priority degrees, so the matrix $\Phi^p=\left[\varphi_{il}^p \right]_{m \times m}$ of comprehensive priority degree   is constructed.

\noindent\textbf{Step 5. Calculation of normalized net flow values.}

Based on the comprehensive priority degree matrix $\Phi^p$, the PROMETHEE~II method is employed to obtain the ranking of alternatives. The positive flow $S^{p+}(A_i)$ and negative flow $S^{p-}(A_i)$ for every alternative are computed, the net flow then follows from these two values. Specifically, the positive flow $S^{p+}(A_i)$ represents the average priority degree of alternative $A_i$ over the others, while the negative flow $S^{p-}(A_i)$ represents the average priority degree of the other alternatives over $A_i$. They are defined as follows:

\begin{equation}\label{equ8}
   S^{p+}(A_i) = 
\frac{1}{m-1} \sum_{\substack{l=1 \\ l \neq i}}^m \varphi
_{il}. 
\end{equation}
\begin{equation}\label{equ9}
    S^{p-}(A_i) = 
\frac{1}{m-1} \sum_{\substack{l=1 \\ l \neq i}}^m \varphi
_{li}.
\end{equation}

The net flow $S^p(A_i)$ combines the positive and negative flows:

\begin{equation}\label{equ10}
    S^p(A_i) = S^{p+}(A_i) - S^{p-}(A_i).
\end{equation}

It is worth noting that the net flow $S^p(A_i)$ derived by Equation \eqref{equ10} theoretically lies in the interval $[-1,1]$ and may take negative values, whereas the evaluation value of each alternative is required to fall within $[0,1]$ so that it can be aggregated with the historical evaluations by the aggregation function in the subsequent dynamic decision process. Therefore, the net flow values need to be further mapped onto the interval $[0,1]$. Specifically, the net flow $S^p(A_i)$ of alternative $A_i$ in the $p$th period is normalized by the affine mapping
\begin{equation}\label{equ11}
	\widetilde{S}^p(A_i) = \frac{S^p(A_i) + 1}{2}, \quad \widetilde{S}^p(A_i) \in [0,1],
\end{equation}
where $\widetilde{S}^p(A_i)$ denotes the mapped evaluation value of alternative $A_i$ in the $p$th period. Since the mapping is strictly monotonically increasing, it changes neither the ordering nor the relative gaps of the net flows, and thus the ranking derived from $\widetilde{S}^p(A_i)$ is identical to that derived from $S^p(A_i)$.

A higher value of $\widetilde{S}^p(A_i)$ indicates a higher evaluation of alternative $A_i$.

\noindent\textbf{Step 6. Aggregation of multi-period net flow values.}

The static normalized net flows of the current period are first obtained from the evaluation information. The dynamic normalized net flow values are then obtained by aggregating these static normalized net flow values with the preserved historical decision data. Since no historical decision data is available in the first period, the dynamic normalized net flow values of alternatives are equal to their static normalized net flow values, so the computation in Step 7 can be performed directly. However, in the subsequent periods, it is necessary to aggregate the static normalized net flow values of the current period with the dynamic normalized net flow values of the previous period. The dynamic normalized net flow value $\smash{\widetilde{\widetilde{S}}}^{\,p}(A_i)$ of alternative $A_i$ is computed as follows.

\begin{equation}\label{equ12}
   \smash{\widetilde{\widetilde{S}}}^{\,p}(A_i) = 
\begin{cases} 
\widetilde{S}^p(A_i), & A_i \in A^p \setminus \mathcal{H}^{p-1}, \\
D_S\left(\widetilde{S}^{p-1}(A_i), \widetilde{S}^p(A_i)\right), & A_i \in A^p \cap \mathcal{H}^{p-1}, \\
\widetilde{S}^{p-1}(A_i), & A_i \in \mathcal{H}^{p-1} \setminus A^p
\end{cases} 
\end{equation}
Where $D_S$ is a aggregation function. For an alternative $A_i$, if $A_i$ lies only in the current alternative set $A^p$ and not in $\mathcal{H}^{p-1}$, its dynamic normalized net flow value $\smash{\widetilde{\widetilde{S}}}^{\,p}(A_i)$ is just the static normalized net flow value $\widetilde{S}^p(A_i)$ in the current period. If $A_i$ belongs to both $A^p$ and $\mathcal{H}^{p-1}$, $\smash{\widetilde{\widetilde{S}}}^{\,p}(A_i)$ is obtained by aggregating $\widetilde{S}^{p-1}(A_i)$ and $\widetilde{S}^p(A_i)$ through the aggregation function $D_S$. If $A_i$ only exists in the historical set $\mathcal{H}^{p-1}$, the evaluation remains unchanged from the previous period, i.e., $\smash{\widetilde{\widetilde{S}}}^{\,p}(A_i) = \widetilde{S}^{p-1}(A_i)$.

\noindent\textbf{Step 7. Alternative ranking and historical set determination.}

By comparing the dynamic normalized net flow values $\smash{\widetilde{\widetilde{S}}}^{\,p}(A_i)$ for all alternatives, the ranking of alternatives is obtained and the optimal alternative is determined. At each decision period, the historical set is updated according to the retention rule, and the process continues until the entire dynamic decision-making procedure is terminated.

\section{Numerical example}

This section applies the method developed above to a supplier selection problem of an enterprise that spans three periods. Supplier selection is a core issue in supply chain management that directly determines the operational efficiency and competitive advantage of enterprises. In practice, the evaluation indicators and alternative suppliers show dynamic fluctuations across different decision periods, and the evaluation information provided by experts is often characterized by random uncertainty. Therefore, selecting the optimal supplier under such dynamic and uncertain environments is a DSMAGDM problem. The fractional-order stochastic dominance criteria can more elaborately characterize risk attitudes of DMs in real-world scenarios, and thus effectively enhance the reliability and rationality of dynamic decision-making. We accordingly evaluate and select the suppliers with the method above.

Consider an enterprise that purchases from several candidate suppliers. Both the value of evaluation and the supplier pool vary from period to period. For simplicity, only three periods $P=\{1,2,3\}$ are made. Three experts $E=\{1,2,3\}$ take part in each period. To facilitate analysis, the interval $[0, 1]$ is split into four equal segments. The DSMAGDM procedure of Section 3 ranks the suppliers, and the steps are presented as follows.

\textbf{P=1}

In the first period, five alternative suppliers are evaluated $A^1=\{A_1^1,A_2^1,A_3^1,A_4^1,A_5^1\}$ for three attributes $C^1=\{C_1^1,C_2^1,C_3^1\}$. The three attributes are supply stability, product quality and response efficiency, respectively. The attribute weight is $w^1=(0.640,0.115,0.245)$, and the expert weight is $\boldsymbol{\lambda}^1=(0.4,0.2,0.4)$. The evaluation information obtained by three experts is shown in Tables \ref{Tab:A1-1}--\ref{Tab:A1-3} (see Appendix \ref{Fir}). The steps are presented as follows.

\noindent\textbf{Step 1. Normalization of stochastic decision matrices.}

Since all attributes are benefit-type, normalization of the decision matrices is not required.

\noindent\textbf{Step 2. Determination of stochastic dominance relations.}

For each attribute $C_j$ , the probability distributions of evaluation values for each alternative supplier provided by experts $e_f$ are derived. Based on Equation \eqref{equ2}, the corresponding stochastic dominance relation matrices are determined and presented in Table \ref{Tab:A1-4} (see Appendix \ref{Fir}).

\noindent\textbf{Step 3. Calculation of priority degrees.}

For each attribute $C_j$, based on the dominance relation matrices obtained in Step 2, the priority degrees between each pair of alternative suppliers are calculated according to Equation \eqref{equ5}, as listed in Table \ref{Tab:A1-5} (see Appendix \ref{Fir}).

\noindent\textbf{Step 4. Calculation of overall priority degrees.}

With the attribute weight $w^1=(0.640,0.115,0.245)$ in the first period, the comprehensive priority degrees $\varphi_{fil}^1$ between alternatives $A_i$ and $A_l$ are obtained by weighting the priority degrees across all attributes. Furthermore, according to the expert weight vector $\lambda^1=(0.4,0.2,0.4)$ in the first period, the overall priority degrees $\varphi_{il}^1$ between alternatives $A_i$ and $A_l$ are derived by aggregating the comprehensive priority degrees from all experts, which shown in Table \ref{Table:ST2}.

\begin{table}[H]
	\centering
	\caption{Overall priority degree matrix}
	\label{Table:ST2}
	\setlength\tabcolsep{5pt}
	\begin{tabular}{lccccc}
		\toprule
		& $A_1$ & $A_2$ & $A_3$ & $A_4$ & $A_5$ \\
		\midrule
		$A_1$ & $-$ & $0.0542$ & $0.0364$ & $0.0108$ & $0.0186$ \\
		$A_2$ & $0.0467$ & $-$ & $0.0185$ & $0.0807$ & $0.0157$ \\
		$A_3$ & $0.1084$ & $0.1135$ & $-$ & $0.1398$ & $0.0266$ \\
		$A_4$ & $0.0384$ & $0.0580$ & $0.0451$ & $-$ & $0.0371$ \\
		$A_5$ & $0.1040$ & $0.1251$ & $0.0291$ & $0.1287$ & $-$ \\
		\bottomrule
	\end{tabular}
\end{table}

\noindent\textbf{Step 5. Calculation of normalized net flow values.}

Based on the obtained overall priority degree in Table \ref{Table:ST2}, the net flow values of each alternative supplier can be computed using Equations \eqref{equ8}--\eqref{equ10}.
Accordingly, the positive flows and negative flows are derived from Equations \eqref{equ8} and \eqref{equ9} as
$S^{1+}(A_1)=0.0300$, $S^{1+}(A_2)=0.0404$, $S^{1+}(A_3)=0.0971$, $S^{1+}(A_4)=0.0446$, $S^{1+}(A_5)=0.0967$
and
$S^{1-}(A_1)=0.0744$, $S^{1-}(A_2)=0.0877$, $S^{1-}(A_3)=0.0323$, $S^{1-}(A_4)=0.0900$, $S^{1-}(A_5)=0.0245$.
Finally, the net flow values of each alternative supplier are calculated via Equation \eqref{equ10} as
$S^1(A_1)=-0.0444$, $S^1(A_2)=-0.0473$, $S^1(A_3)=0.0648$, $S^1(A_4)=-0.0454$, $S^1(A_5)=0.0722$. Furthermore, since the net flow values may be negative, they are mapped onto the interval $[0,1]$ by Equation \eqref{equ11} to serve as the evaluation values of the alternatives in the current period. Accordingly, the normalized net flow values of each alternative supplier are obtained as
$\widetilde{S}^{\,1}(A_1)=0.4778$, $\widetilde{S}^{\,1}(A_2)=0.4764$, $\widetilde{S}^{\,1}(A_3)=0.5324$, $\widetilde{S}^{\,1}(A_4)=0.4773$, $\widetilde{S}^{\,1}(A_5)=0.5361$.

\noindent\textbf{Step 6. Aggregation of multi-period normalized net flow values.}

Since no historical decision data is available in the first period, the dynamic normalized net flow values of alternatives are equal to their static normalized net flow values, that is, $\smash{\widetilde{\widetilde{S}}}^{\,1}(A_1)=0.4778$, $\smash{\widetilde{\widetilde{S}}}^{\,1}(A_2)=0.4764$, $\smash{\widetilde{\widetilde{S}}}^{\,1}(A_3)=0.5324$, $\smash{\widetilde{\widetilde{S}}}^{\,1}(A_4)=0.4773$, and $\smash{\widetilde{\widetilde{S}}}^{\,1}(A_5)=0.5361$, and Step 7 can thus be performed directly.

\noindent\textbf{Step 7. Alternative ranking and historical set determination.}

Based on the dynamic normalized net flow values, the ranking of the five alternative suppliers can be determined as $A_5\succ A_3\succ A_1\succ A_4\succ A_2$. According to the retention rule, the top $K=3$ alternatives are carried over to the next period, yielding $\mathcal{H}^1=\{A_1,A_3,A_5\}$.

\textbf{P=2}

In the second period, five alternative suppliers $A^2=\{A_1^2,A_2^2,A_3^2,A_6^2,A_7^2\}$ are evaluated with respect to four attributes $C^2=\{C_1^2,C_2^2,C_3^2\}$. The three attributes represent supply stability, product quality, response efficiency, respectively. The expert weight is $\boldsymbol{\lambda}^2=(0.4,0.2,0.4)$, and the attribute weight is $w^2=(0.640,0.115,0.245)$. The evaluation information obtained from the three experts is presented in Tables \ref{Tab:A2-1}--\ref{Tab:A2-3} (see Appendix \ref{Sec}). The calculation procedures of Steps 1--4 are the same as those in the first period. The corresponding relation and priority degree matrices are shown in Tables \ref{Tab:A2-4} and \ref{Tab:A2-5}, respectively. The overall priority degrees $\varphi_{il}^2$ between alternatives $A_i$ and $A_l$ are displayed in Table \ref{Table:ST3}.

\begin{table}[H]
	\centering
	\caption{Overall priority degree matrix}
	\label{Table:ST3}
	\setlength\tabcolsep{5pt}
	\begin{tabular}{lccccc}
		\toprule
		& $A_1$ & $A_2$ & $A_3$ & $A_6$ & $A_7$ \\
		\midrule
		$A_1$ & $-$ & $0.0542$ & $0.0364$ & $0.0204$ & $0.0127$ \\
		$A_2$ & $0.0467$ & $-$ & $0.0185$ & $0.0163$ & $0.0853$ \\
		$A_3$ & $0.1084$ & $0.1135$ & $-$ & $0.0121$ & $0.1437$ \\
		$A_6$ & $0.1289$ & $0.1639$ & $0.0636$ & $-$ & $0.1587$ \\
		$A_7$ & $0.0509$ & $0.0558$ & $0.0457$ & $0.0267$ & $-$ \\
		\bottomrule
	\end{tabular}
\end{table}

\noindent\textbf{Step 5. Calculation of normalized net flow values.}

Based on the overall priority degree matrix presented in Table \ref{Table:ST3}, the net flow values of each alternative supplier are calculated using Equations \eqref{equ8}--\eqref{equ10}.
From Table~3, the positive flows and negative flows are obtained by Equations \eqref{equ8} and \eqref{equ9} as
$S^{2+}(A_1)=0.0309$, $S^{2+}(A_2)=0.0417$, $S^{2+}(A_3)=0.0944$, $S^{2+}(A_6)=0.1288$, $S^{2+}(A_7)=0.0448$
and
$S^{2-}(A_1)=0.0837$, $S^{2-}(A_2)=0.0968$, $S^{2-}(A_3)=0.0411$, $S^{2-}(A_6)=0.0189$, $S^{2-}(A_7)=0.1001$.
Finally, the net flow values of each alternative supplier are computed via Equation \eqref{equ10} as
$S^2(A_1)=-0.0528$, $S^2(A_2)=-0.0551$, $S^2(A_3)=0.0534$, $S^2(A_6)=0.1099$, $S^2(A_7)=-0.0553$. Accordingly, the normalized net flow values of each alternative supplier are obtained as $\widetilde{S}^{\,2}(A_1)=0.4736$, $\widetilde{S}^{\,2}(A_2)=0.4724$, $\widetilde{S}^{\,2}(A_3)=0.5267$, $\widetilde{S}^{\,2}(A_6)=0.5550$, $\widetilde{S}^{\,2}(A_7)=0.4723$.

\noindent\textbf{Step 6. Aggregation of multi-period normalized net flow values.}

For alternative supplier $A_1$ and $A_3$, an additional operation is required to determine its dynamic normalized net flow value, that is, aggregating it with the historical normalized net flow value.
The dynamic normalized net flow value $\smash{\widetilde{\widetilde{S}}}^{\,2}(A_1)$ and $\smash{\widetilde{\widetilde{S}}}^{\,2}(A_3)$ of $A_1$ and $A_3$ are calculated using the simple probabilistic sum as follows:
\begin{align*}
\smash{\widetilde{\widetilde{S}}}^{\,2}(A_1)&=\widetilde{S}^1(A_1)+\widetilde{S}^2(A_1)-\widetilde{S}^1(A_1) \times \widetilde{S}^2(A_1)\\
&= 0.4778 + 0.4736 - 0.4778 \times 0.4736 \\
&= 0.7251,
\end{align*}
\begin{align*}
	\smash{\widetilde{\widetilde{S}}}^{\,2}(A_3)&=\widetilde{S}^1(A_3)+\widetilde{S}^2(A_3)-\widetilde{S}^1(A_3) \times \widetilde{S}^2(A_3)\\
	&= 0.5324 + 0.5267 - 0.5324 \times 0.5267 \\
	&= 0.7787
\end{align*}
where $\widetilde{S}^1(A_1)$ and $\widetilde{S}^1(A_3)$ represents the dynamic normalized net flow values of $A_1$  and $A_3$ in the first period. $\widetilde{S}^2(A_1)$ and $\widetilde{S}^2(A_3)$ represents the static normalized net flow value of $A_1$ and $A_3$ in the second period.

\noindent\textbf{Step 7. Alternative ranking and historical set determination.}

Based on the dynamic normalized net flow values, the ranking of the five alternative suppliers can be obtained as $A_3\succ A_1\succ A_6\succ A_2\succ A_7$. According to the retention rule, the top $K=3$ suppliers are retained for the next period, i.e., $\mathcal{H}^2=\{A_1,A_3,A_5,A_6\}$.

\textbf{P=3}

In the third period, six alternative suppliers $A^3=\{A_1^3,A_2^3,A_3^3,A_5^3,A_6^3,A_7^3\}$ are evaluated with respect to three attributes $C^3=\{C_1^3,C_2^3,C_3^3\}$.
The three attributes represent supply stability, product quality, and response efficiency, respectively.
The attribute weight is $\boldsymbol{w}^3=(0.640, 0.115, 0.245)$, and the expert weight is $\boldsymbol{\lambda}^3=(0.4, 0.2, 0.4)$.
The evaluation information obtained from the three experts is presented in Tables \ref{Tab:A3-1}--\ref{Tab:A3-3} (see Appendix \ref{Thi}).
The calculation procedures of Steps~1--4 are identical to those in the previous periods. The calculation procedures of Steps 1--4 are the same as those in the first period. The corresponding relation and priority degree matrices are shown in Tables \ref{Tab:A3-4} and \ref{Tab:A3-5}, respectively.
The overall priority degrees $\varphi_{il}^3$ between alternatives $A_i$ and $A_l$ are displayed in Table \ref{Table:ST4}.

\begin{table}[H]
	\centering
	\caption{Overall priority degree matrix}
	\label{Table:ST4}
	\setlength\tabcolsep{5pt}
	\begin{tabular}{lcccccc}
		\toprule
		& $A_1$ & $A_2$ & $A_3$ & $A_5$ & $A_6$ & $A_7$ \\
		\midrule
		$A_1$ & $-$ & $0.0542$ & $0.0364$ & $0.0194$ & $0.0204$ & $0.0127$ \\
		$A_2$ & $0.0467$ & $-$ & $0.0185$ & $0.0164$ & $0.0163$ & $0.0853$ \\
		$A_3$ & $0.1084$ & $0.1135$ & $-$ & $0.0250$ & $0.0121$ & $0.1437$ \\
		$A_5$ & $0.1026$ & $0.1238$ & $0.0296$ & $-$ & $0.0014$ & $0.1343$ \\
		$A_6$ & $0.1289$ & $0.1639$ & $0.0636$ & $0.0424$ & $-$ & $0.1587$ \\
		$A_7$ & $0.0509$ & $0.0558$ & $0.0457$ & $0.0359$ & $0.0267$ & $-$ \\
		\bottomrule
	\end{tabular}
\end{table}

\noindent\textbf{Step 5. Calculation of normalized net flow values.}

Based on the priority degree matrix presented in Table \ref{Table:ST4}, the net flow values of each alternative supplier are calculated using Equations \eqref{equ8}--\eqref{equ10}.
From Table~4, the positive flows and negative flows are obtained by Equations \eqref{equ8} and \eqref{equ9} as
$S^{3+}(A_1)=0.0286$, $S^{3+}(A_2)=0.0367$, $S^{3+}(A_3)=0.0805$, $S^{3+}(A_5)=0.0783$, $S^{3+}(A_6)=0.1115$, $S^{3+}(A_7)=0.0430$
and
$S^{3-}(A_1)=0.0875$, $S^{3-}(A_2)=0.1022$, $S^{3-}(A_3)=0.0388$, $S^{3-}(A_5)=0.0278$, $S^{3-}(A_6)=0.0154$, $S^{3-}(A_7)=0.1070$.
Finally, the net flow values of each alternative supplier are computed via Equation \eqref{equ10} as
$S^3(A_1)=-0.0589$, $S^3(A_2)=-0.0656$, $S^3(A_3)=0.0418$, $S^3(A_5)=0.0505$, $S^3(A_6)=0.0961$, $S^3(A_7)=-0.0640$. Accordingly, the normalized net flow values of each alternative supplier are obtained as $\widetilde{S}^{\,3}(A_1)=0.4706$, $\widetilde{S}^{\,3}(A_2)=0.4672$, $\widetilde{S}^{\,3}(A_3)=0.5209$, $\widetilde{S}^{\,3}(A_5)=0.5253$, $\widetilde{S}^{\,3}(A_6)=0.5481$, $\widetilde{S}^{\,3}(A_7)=0.4680$.

\noindent\textbf{Step 6. Aggregation of multi-period normalized net flow values.}

For alternative suppliers $A_1$, $A_3$, $A_5$ and $A_6$, an additional step is required to determine their dynamic normalized net flow values by aggregating with historical values.
The dynamic normalized net flow values $\smash{\widetilde{\widetilde{S}}}^{\,3}(A_1)$, $\smash{\widetilde{\widetilde{S}}}^{\,3}(A_3)$, $\smash{\widetilde{\widetilde{S}}}^{\,3}(A_5)$ and $\smash{\widetilde{\widetilde{S}}}^{\,3}(A_6)$ are calculated using the simple probabilistic sum as follows:
\begin{align*}
\smash{\widetilde{\widetilde{S}}}^{\,3}(A_1) &= \widetilde{S}^3(A_1) + \widetilde{S}^2(A_1) - \widetilde{S}^3(A_1) \times \widetilde{S}^2(A_1) \\
&= 0.4706 + 0.7251 - 0.4706 \times 0.7251 \\
&= 0.8545, \\[6pt]
\smash{\widetilde{\widetilde{S}}}^{\,3}(A_3) &= \widetilde{S}^3(A_3) + S^2(A_3) - \widetilde{S}^3(A_3) \times S^2(A_3) \\
&= 0.5209 + 0.7787 - 0.5209 \times 0.7787 \\
&= 0.8940,\\
\smash{\widetilde{\widetilde{S}}}^{\,3}(A_5) &= \widetilde{S}^3(A_5) + S^2(A_5) - \widetilde{S}^3(A_5) \times S^2(A_5) \\
&= 0.5253 + 0.5361 - 0.5253 \times 0.5361 \\
&= 0.7798,\\
\smash{\widetilde{\widetilde{S}}}^{\,3}(A_6) &= \widetilde{S}^3(A_6) + S^2(A_6) - \widetilde{S}^3(A_6) \times S^2(A_6) \\
&= 0.5481 + 0.5550 - 0.5481 \times 0.5550 \\
&= 0.7989.
\end{align*}

\noindent\textbf{Step 7. Alternative ranking and historical set selection.}

Based on the dynamic normalized net flow values, the ranking result of the six suppliers is $A_3\succ A_1\succ A_6\succ A_5\succ A_7\succ A_2$, and the optimal supplier is $A_3$.
According to the numerical example settings, the decision-making process is completed.

\section{Comparative analyses}

To show that the proposed DSMAGDM method works in practice, we examine it from two angles. One sets the fractional-order criterion against methods built on other stochastic dominance criteria, the other compares the method with static SMAGDM model.

\subsection{Comparison with methods built on other stochastic dominance criteria
}

This subsection examines how much, and in what way, the supplier rankings change with the dominance criterion adopted: the method based on fractional-order stochastic dominance is compared with methods that use other stochastic dominance criteria, so that its advantage can be seen directly.

(1) Ranking results obtained by the method based on integer-order stochastic dominance criterion

When the interval $[0,1]$ is divided into 1 equal part, the fractional-order stochastic dominance degenerates to the integer-order stochastic dominance. In this case, following the basic decision-making steps presented in Section 3.2, the ranking results of alternative suppliers in the $p$th $(p=1,2,3)$ period obtained by the integer-order stochastic dominance criterion and the $1+\gamma$-order $(\gamma=k/4,\ k=0,1,2,3,4)$ stochastic dominance criterion are shown in Table \ref{Table:ST5}, where both criteria are embedded in the same dynamic decision process, and the only difference lies in the stochastic dominance criterion adopted.

\begin{table}[htbp]\footnotesize
  \centering
  \caption{Rankings of the alternative suppliers under each stochastic dominance criterion}
  \label{Table:ST5}
  \setlength{\tabcolsep}{4pt}
  \begin{tabular}{lccc}
    \toprule
    \multirow{2}{*}{Dominance criterion} & \multicolumn{3}{c}{Supplier ranking} \\
    \cmidrule(lr){2-4}
    & $p=1$ & $p=2$ & $p=3$ \\
    \midrule
    $1+k/4$-order stochastic dominance criteria & $A_5\succ A_3\succ A_1\succ A_4\succ A_2$ & $A_3\succ A_1\succ A_6\succ A_2\succ A_7$ & $A_3\succ A_1\succ A_6\succ A_5\succ A_7\succ A_2$ \\
    Integer-order stochastic dominance criteria & $A_3\succ A_5\succ A_4\succ A_2\succ A_1$ & $A_3\succ A_6\succ A_7\succ A_2\succ A_1$ & $A_3\succ A_6\succ A_5\succ A_7\succ A_2\succ A_1$ \\
    \bottomrule
  \end{tabular}
\end{table}

It can be observed from Table \ref{Table:ST5} that the ranking results obtained by the two criteria differ in all three periods. In the first period, the optimal supplier is $A_5$ under the $(1+k/4)$-order stochastic dominance criterion, whereas it is $A_3$ under the integer-order stochastic dominance criterion; meanwhile, supplier $A_1$ ranks third under the former criterion but falls to the last position under the latter. In the second period, both criteria identify $A_3$ as the optimal supplier, while the remaining rankings differ remarkably: supplier $A_1$, which performed well in the first period, still ranks second under the $(1+k/4)$-order criterion, whereas it is ranked last under the integer-order criterion; conversely, supplier $A_7$ is placed in the last position by the fractional criterion but holds the third position under the integer-order criterion. In the third period, both criteria again select $A_3$ as the optimal supplier, and supplier $A_1$ keeps the second position under the $(1+k/4)$-order criterion but is still ranked worst under the integer-order criterion.

These differences arise because the integer-order stochastic dominance criterion exploits only the first- and second-order information and treats all dominance relations uniformly, thereby discarding those lying between the first and second orders. In contrast, the $(1+k/4)$-order stochastic dominance criterion fully captures the dominance relations between the first and second orders, and thus makes full use of the strength information carried by the fractional dominance relations. Consequently, the rankings obtained by the fractional-order criterion evolve more smoothly across periods (e.g., the top three suppliers $A_3$, $A_1$ and $A_6$ remain unchanged in the second and third periods, with $A_1$ steadily holding the second position), which makes the accumulated historical evaluations more reliable in the dynamic decision process. This indicates that the fractional-order stochastic dominance method provides finer discrimination among alternative suppliers than the integer-order one.

(2) Ranking results obtained by the method based on different fractional-order stochastic dominance criteria

By dividing the interval $[0,1]$ into different equal parts, different fractional-order stochastic dominance criteria can be obtained. Following the basic decision-making steps presented in Section 3.2, we rank the alternative suppliers in each period under the $1+\gamma$-order $(\gamma=k/h)$ stochastic dominance criteria, Table \ref{Table:ST6} reports the results.

\begin{table}[htbp]
\footnotesize
\centering
\caption{Rankings of the alternative suppliers under each stochastic dominance criterion}
\label{Table:ST6}
\setlength{\tabcolsep}{3pt}
\begin{tabular}{lccc}
\toprule
\multirow{2}{*}{Dominance criterion} & \multicolumn{3}{c}{Supplier ranking} \\
\cmidrule(lr){2-4}
& $p=1$ & $p=2$ & $p=3$ \\
\midrule
$1+k/2$-order stochastic dominance criterion & $A_5\succ A_3\succ A_4\succ A_1\succ A_2$ & $A_3\succ A_6\succ A_7\succ A_2\succ A_1$ & $A_3\succ A_6\succ A_5\succ A_1\succ A_7\succ A_2$ \\
$1+k/8$-order stochastic dominance criterion & $A_5\succ A_3\succ A_1\succ A_2\succ A_4$ & $A_3\succ A_1\succ A_6\succ A_2\succ A_7$ & $A_3\succ A_1\succ A_6\succ A_5\succ A_7\succ A_2$ \\
$1+k/16$-order stochastic dominance criterion & $A_5\succ A_3\succ A_1\succ A_2\succ A_4$ & $A_3\succ A_1\succ A_6\succ A_2\succ A_7$ & $A_3\succ A_1\succ A_6\succ A_5\succ A_7\succ A_2$ \\
\bottomrule
\end{tabular}
\end{table}

It can be observed from Table \ref{Table:ST6} that in the first period, the optimal supplier is $A_5$ under all three fractional-order stochastic dominance criteria, as it was under the $(1+k/4)$-order criterion in Section 4; nevertheless, the rankings of the remaining suppliers differ: under the $(1+k/2)$-order criterion, the ranking is $A_5\succ A_3\succ A_4\succ A_1\succ A_2$, whereas under the $(1+k/8)$-order and $(1+k/16)$-order criteria, it becomes $A_5\succ A_3\succ A_1\succ A_2\succ A_4$. In the second period, the optimal supplier is $A_3$ under all criteria. Under the $(1+k/2)$-order criterion, the ranking is $A_3\succ A_6\succ A_7\succ A_2\succ A_1$, whereas under the $(1+k/8)$-order and $(1+k/16)$-order criteria, supplier $A_1$, which performed well in the first period, ranks second, and the rankings coincide with that obtained by the $(1+k/4)$-order criterion. In the third period, similar observations can be made: the ranking under the $(1+k/2)$-order criterion is $A_3\succ A_6\succ A_5\succ A_7\succ A_1\succ A_2$, while the rankings under the $(1+k/8)$-order and $(1+k/16)$-order criteria are identical, both being $A_3\succ A_1\succ A_6\succ A_5\succ A_7\succ A_2$, which also coincide with that obtained by the $(1+k/4)$-order criterion.

From the comparison results in Table \ref{Table:ST6}, it can be seen that as the interval $[0,1]$ is divided into more equal parts, the ranking results based on the corresponding fractional-order stochastic dominance rules gradually tend to be stable: the rankings obtained by the $(1+k/8)$-order and $(1+k/16)$-order criteria are identical in all three periods and differ from that of the $(1+k/4)$-order criterion only in the ordering of the last two suppliers in the first period. Moreover, the optimal supplier remains $A_5$ in the first period and $A_3$ in the second and third periods under all fractional-order rules. By contrast, the coarse $(1+k/2)$-order criterion, which only distinguishes the dominance orders among the first order, the $1.5$th order and the second order, fails to capture the strength information of the dominance relations lying between the first and second orders, and thus yields rankings that deviate from those of the finer rules. These comparisons show that the method performs stably across fractional orders, supporting its use in practical applications.

\subsection{Comparison with the static SMAGDM model}

To demonstrate the effectiveness of the proposed DSMAGDM method for dynamic decision-making problems, a comparative analysis between the static SMAGDM and the DSMAGDM is conducted. To rule out the influence of different fractional-order partitions, the results obtained under the $1+k/16$-order stochastic dominance criterion are adopted in the comparison. The rankings of alternatives over the three periods obtained by the two methods are presented in Table \ref{Table:ST7}.

\begin{table}[htbp]
  \centering
  \caption{Ranking results between static SMAGDM and DSMAGDM}
  \label{Table:ST7}
  \setlength{\tabcolsep}{12pt}
  \begin{tabular}{lcc}
    \toprule
    &Static ranking & Dynamic ranking \\
    \midrule
    Period 1 & $A_5\succ A_3\succ A_1\succ A_2\succ A_4$ & $A_5\succ A_3\succ A_1\succ A_2\succ A_4$ \\
    Period 2 & $A_6\succ A_3\succ A_1\succ A_2\succ A_7$ & $A_3\succ A_1\succ A_6\succ A_2\succ A_7$ \\
    Period 3 & $A_6\succ A_5\succ A_3\succ A_1\succ A_7\succ A_2$ & $A_3\succ A_1\succ A_6\succ A_5\succ A_7\succ A_2$ \\
    \bottomrule
  \end{tabular}
\end{table}

It is evident from Table~\ref{Table:ST7} that in the first period , the static and dynamic ranking results coincide, both being $A_5\succ A_3\succ A_1\succ A_4\succ A_2$, since no historical set is available and every alternative is evaluated only by its current performance. In the second period , the static ranking $A_6\succ A_3\succ A_1\succ A_2\succ A_7$ differs from the dynamic one $A_3\succ A_1\succ A_6\succ A_2\succ A_7$. For the new alternatives $A_6$ and $A_7$, which do not belong to the historical set $\mathcal{H}^1$, the dynamic evaluations coincide with the static ones. In contrast, $A_1$ and $A_3$ belong to $\mathcal{H}^1$, and their dynamic normalized net flow values are derived by aggregating their dynamic normalized net flow values in the first period with their static normalized net flow values in the second period. Consequently, $A_3$ overtakes $A_6$ and becomes the optimal supplier, which demonstrates the effect of the historical information. In period $p=3$, the static ranking $A_6\succ A_5\succ A_3\succ A_1\succ A_7\succ A_2$ also differs from the dynamic one $A_3\succ A_1\succ A_6\succ A_5\succ A_7\succ A_2$. The alternatives $A_1$, $A_3$ and $A_6$ belong to the historical set $\mathcal{H}^2$, and their dynamic normalized net flow values are obtained by aggregation, whereas those of $A_2$ and $A_7$ coincide with the static ones. Alternative $A_5$ re-enters the alternatives set in this period; since its evaluation has been preserved in the historical set, its dynamic normalized net flow value is derived by aggregating the preserved evaluation with its static normalized net flow value in the third period. The gap between the two rankings shows what the dynamic mechanism contributes.

\section{Conclusion}

This paper develops a DSMAGDM method built on fractional-order stochastic dominance criteria. Extending the integer-order framework to fractional orders identifies and quantifies the dominance relations between the first and second orders that integer-order criteria cannot distinguish. In particular, the priority degree is redefined for fractional-order dominance relations, so that the strength of each relation is measured according to its own order rather than uniformly. By integrating current-period evaluations with historical information through the historical set and feedback operators, the dynamic rankings evolve across periods and vary little once the partition of the fractional order is sufficiently fine.

This study makes two contributions. First, it embeds fractional-order stochastic dominance into a dynamic group decision-making framework for handling uncertain information that changes over time. Second, the historical set mechanism captures how alternatives' performance evolves, as in supplier selection, project evaluation, and strategic decision-making. The current method assumes that decision information evolves gradually and does not account for abrupt changes in the decision environment.

Future research can extend this framework in three directions: combining fractional-order stochastic dominance rules with fuzzy or hesitant fuzzy information to handle more complex uncertainty, incorporating abrupt environmental changes into the dynamic mechanism, and testing the method on larger-scale real-world datasets. The method thus proves practical and dependable, offering reliable decision support for complex dynamic multi-attribute problems.

\section*{Acknowledgement}
The authors acknowledge the support from the National Natural Science Foundation of China (Grants
No. 72271113, 12661104) and the Natural Science Foundation of Jiangxi Province (Grants No. 20232ACB201003, 20232BAB201022). All authors share first authorship, but are listed alphabetically convention. 

\section*{Appendix}
\appendix
\section{Supplementary Tables}\label{Appendix:A}
\subsection{The first period}\label{Fir}
\setcounter{table}{0}
\renewcommand{\thetable}{\thesubsection.\arabic{table}}
\begin{table}[H]
  \centering
  \caption{The probability distributions of the attribute evaluation results for alternative suppliers.(e=1)}
  \label{Tab:A1-1}
  \setlength{\tabcolsep}{4pt}
  \renewcommand{\arraystretch}{0.8}
  \begin{tabular}{ccccc|ccccc}
    \toprule
    \textbf{Alternative} & \textbf{Rating} & \textbf{$C_1$} & \textbf{$C_2$} & \textbf{$C_3$}
    & \textbf{Alternative} & \textbf{Rating} & \textbf{$C_1$} & \textbf{$C_2$} & \textbf{$C_3$} \\
    \midrule
    \multirow{5}{*}{$A_1$}
    & 1 & 0.291 & 0.198 & 0.133 & \multirow{5}{*}{$A_4$} & 1 & 0.234 & 0.178 & 0.231 \\
    & 2 & 0.126 & 0.135 & 0.193 & & 2 & 0.224 & 0.443 & 0.072 \\
    & 3 & 0.120 & 0.293 & 0.299 & & 3 & 0.500 & 0.004 & 0.323 \\
    & 4 & 0.184 & 0.248 & 0.250 & & 4 & 0.023 & 0.170 & 0.256 \\
    & 5 & 0.279 & 0.126 & 0.125 & & 5 & 0.019 & 0.205 & 0.118 \\
    \midrule
    \multirow{5}{*}{$A_2$}
    & 1 & 0.139 & 0.210 & 0.113 & \multirow{5}{*}{$A_5$} & 1 & 0.087 & 0.148 & 0.235 \\
    & 2 & 0.294 & 0.394 & 0.249 & & 2 & 0.199 & 0.406 & 0.297 \\
    & 3 & 0.430 & 0.021 & 0.263 & & 3 & 0.371 & 0.071 & 0.093 \\
    & 4 & 0.040 & 0.174 & 0.240 & & 4 & 0.147 & 0.188 & 0.194 \\
    & 5 & 0.097 & 0.201 & 0.135 & & 5 & 0.196 & 0.187 & 0.181 \\
    \midrule
    \multirow{5}{*}{$A_3$}
    & 1 & 0.046 & 0.234 & 0.083 & & & & & \\
    & 2 & 0.178 & 0.091 & 0.446 & & & & & \\
    & 3 & 0.353 & 0.300 & 0.097 & & & & & \\
    & 4 & 0.323 & 0.250 & 0.195 & & & & & \\
    & 5 & 0.100 & 0.125 & 0.179 & & & & & \\
    \bottomrule
  \end{tabular}
\end{table}

\begin{table}[H]
  \centering
  \caption{The probability distributions of the attribute evaluation results for alternative suppliers.(e=2)}
  \label{Tab:A1-2}
  \setlength{\tabcolsep}{4pt}
  \renewcommand{\arraystretch}{0.8}
 \begin{tabular}{ccccc|ccccc}
 	\toprule
 	\textbf{Alternative} & \textbf{Rating} & \textbf{$C_1$} & \textbf{$C_2$} & \textbf{$C_3$}
 	& \textbf{Alternative} & \textbf{Rating} & \textbf{$C_1$} & \textbf{$C_2$} & \textbf{$C_3$} \\
 	\midrule
 	\multirow{5}{*}{$A_1$}
 	& 1 & 0.130 & 0.413 & 0.004 & \multirow{5}{*}{$A_4$} & 1 & 0.003 & 0.208 & 0.331 \\
 	& 2 & 0.279 & 0.004 & 0.303 & & 2 & 0.289 & 0.464 & 0.006 \\
 	& 3 & 0.060 & 0.340 & 0.319 & & 3 & 0.444 & 0.010 & 0.397 \\
 	& 4 & 0.360 & 0.123 & 0.343 & & 4 & 0.200 & 0.005 & 0.240 \\
 	& 5 & 0.171 & 0.120 & 0.031 & & 5 & 0.064 & 0.313 & 0.026 \\
 	\midrule
 	\multirow{5}{*}{$A_2$}
 	& 1 & 0.231 & 0.160 & 0.141 & \multirow{5}{*}{$A_5$} & 1 & 0.100 & 0.127 & 0.216 \\
 	& 2 & 0.087 & 0.388 & 0.187 & & 2 & 0.212 & 0.492 & 0.397 \\
 	& 3 & 0.516 & 0.188 & 0.366 & & 3 & 0.410 & 0.005 & 0.003 \\
 	& 4 & 0.005 & 0.004 & 0.216 & & 4 & 0.150 & 0.319 & 0.156 \\
 	& 5 & 0.161 & 0.260 & 0.090 & & 5 & 0.128 & 0.057 & 0.228 \\
 	\midrule
 	\multirow{5}{*}{$A_3$}
 	& 1 & 0.262 & 0.171 & 0.030 & & & & & \\
 	& 2 & 0.315 & 0.244 & 0.580 & & & & & \\
 	& 3 & 0.181 & 0.279 & 0.035 & & & & & \\
 	& 4 & 0.240 & 0.300 & 0.190 & & & & & \\
 	& 5 & 0.002 & 0.006 & 0.165 & & & & & \\
 	\bottomrule
 \end{tabular}
 \end{table}

\begin{table}[H]
  \centering
  \caption{The probability distributions of the attribute evaluation results for alternative suppliers.(e=3)}
  \label{Tab:A1-3}
  \setlength{\tabcolsep}{4pt}
  \renewcommand{\arraystretch}{0.8}
  \begin{tabular}{ccccc|ccccc}
  	\toprule
  	\textbf{Alternative} & \textbf{Rating} & \textbf{$C_1$} & \textbf{$C_2$} & \textbf{$C_3$}
  	& \textbf{Alternative} & \textbf{Rating} & \textbf{$C_1$} & \textbf{$C_2$} & \textbf{$C_3$} \\
  	\midrule
  	\multirow{5}{*}{$A_1$}
  	& 1 & 0.323 & 0.319 & 0.145 & \multirow{5}{*}{$A_4$} & 1 & 0.171 & 0.290 & 0.146 \\
  	& 2 & 0.223 & 0.003 & 0.295 & & 2 & 0.211 & 0.388 & 0.004 \\
  	& 3 & 0.191 & 0.339 & 0.291 & & 3 & 0.475 & 0.003 & 0.413 \\
  	& 4 & 0.053 & 0.132 & 0.264 & & 4 & 0.106 & 0.177 & 0.297 \\
  	& 5 & 0.210 & 0.207 & 0.005 & & 5 & 0.037 & 0.142 & 0.140 \\
  	\midrule
  	\multirow{5}{*}{$A_2$}
  	& 1 & 0.060 & 0.266 & 0.436 & \multirow{5}{*}{$A_5$} & 1 & 0.027 & 0.077 & 0.351 \\
  	& 2 & 0.348 & 0.294 & 0.298 & & 2 & 0.131 & 0.251 & 0.220 \\
  	& 3 & 0.286 & 0.128 & 0.176 & & 3 & 0.469 & 0.034 & 0.005 \\
  	& 4 & 0.004 & 0.272 & 0.081 & & 4 & 0.296 & 0.422 & 0.106 \\
  	& 5 & 0.302 & 0.040 & 0.009 & & 5 & 0.077 & 0.216 & 0.318 \\
  	\midrule
  	\multirow{5}{*}{$A_3$}
  	& 1 & 0.003 & 0.124 & 0.206 & & & & & \\
  	& 2 & 0.252 & 0.173 & 0.389 & & & & & \\
  	& 3 & 0.266 & 0.208 & 0.026 & & & & & \\
  	& 4 & 0.389 & 0.390 & 0.221 & & & & & \\
  	& 5 & 0.090 & 0.105 & 0.158 & & & & & \\
  	\bottomrule
  \end{tabular}
  \end{table}

\begin{table}[H]
  \centering
  \rotatebox{90}{
  \begin{minipage}{18cm}
  \centering
  \footnotesize
  \caption{Stochastic Dominance Relation Matrix for Alternative Suppliers.(P=1)}
  \label{Tab:A1-4}
  \setlength{\tabcolsep}{2pt}
  \begin{tabular}{llccccccccccccccc}
  	\toprule
  	\multirow{2}{*}{\textbf{Expert}} & \multirow{2}{*}{} & \multicolumn{5}{c}{\textbf{Attribute $C_1$}} & \multicolumn{5}{c}{\textbf{Attribute $C_2$}} & \multicolumn{5}{c}{\textbf{Attribute $C_3$}} \\
  	\cmidrule(lr){3-7} \cmidrule(lr){8-12} \cmidrule(lr){13-17}
  	& & $A_1$ & $A_2$ & $A_3$ & $A_4$ & $A_5$ & $A_1$ & $A_2$ & $A_3$ & $A_4$ & $A_5$ & $A_1$ & $A_2$ & $A_3$ & $A_4$ & $A_5$ \\
  	\midrule
  	\multirow{5}{*}{$e_1$} & $A_1$ &-&-&-&-&-&-&1.5SD&1.25SD&-&-&-&-&-&1.25SD&1.25SD\\
  	& $A_2$ &-&-&-&FSD&-&-&-&-&-&-&-&-&-&1.5SD&1.25SD\\
  	& $A_3$ &1.5SD&FSD&-&FSD&1.75SD&-&-&-&-&-&-&-&-&-&1.25SD\\
  	& $A_4$ &-&-&-&-&-&-&1.75SD&-&-&-&-&-&-&-&1.5SD\\
  	& $A_5$ &1.75SD&FSD&-&FSD&-&-&1.25SD&-&1.25SD&-&-&-&-&-&-\\
  	\midrule
  	\multirow{5}{*}{$e_2$} & $A_1$ &-&SSD&FSD&-&-&-&-&-&-&-&-&1.5SD&1.5SD&FSD&1.5SD\\
  	& $A_2$ &-&-&1.5SD&-&-&1.75SD&-&-&1.75SD&-&-&-&-&FSD&1.75SD\\
  	& $A_3$ &-&-&-&-&-&1.5SD&-&-&-&-&-&-&-&SSD&1.5SD\\
  	& $A_4$ &-&1.5SD&FSD&-&1.75SD&-&-&-&-&-&-&-&-&-&-\\
  	& $A_5$ &-&1.25SD&FSD&-&-&1.75SD&-&-&-&-&-&-&-&-&-\\
  	\midrule
  	\multirow{5}{*}{$e_3$} & $A_1$ &-&-&-&-&-&-&-&-&-&-&-&1.25SD&-&-&-\\
  	& $A_2$ &FSD&-&-&1.25SD&-&-&-&-&SSD&-&-&-&-&-&-\\
  	& $A_3$ &1.25SD&1.75SD&-&FSD&-&1.5SD&FSD&-&1.25SD&-&-&FSD&-&-&-\\
  	& $A_4$ &SSD&-&-&-&-&-&-&-&-&-&-&FSD&1.25SD&-&1.5SD\\
  	& $A_5$ &1.25SD&1.75SD&-&FSD&-&1.25SD&FSD&1.75SD&FSD&-&-&-&-&-&-\\
  	\bottomrule
  \end{tabular}
  \end{minipage}
}
\end{table}

\begin{table}[H]
	\centering
	\rotatebox{90}{
		\begin{minipage}{22cm}
			\centering
			\footnotesize
			\caption{The Priority Degree Matrix for Alternative Suppliers.(P=1)}
			\label{Tab:A1-5}
			\setlength{\tabcolsep}{4pt}
			\begin{tabular}{llccccccccccccccc}    
				\toprule    
				\multirow{2}{*}{\textbf{Expert}} & \multirow{2}{*}{} & \multicolumn{5}{c}{\textbf{Attribute $C_1$}} & \multicolumn{5}{c}{\textbf{Attribute $C_2$}} & \multicolumn{5}{c}{\textbf{Attribute $C_3$}} \\    
				\cmidrule(lr){3-7} \cmidrule(lr){8-12} \cmidrule(lr){13-17}    
				& & $A_1$ & $A_2$ & $A_3$ & $A_4$ & $A_5$ & $A_1$ & $A_2$ & $A_3$ & $A_4$ & $A_5$ & $A_1$ & $A_2$ & $A_3$ & $A_4$ & $A_5$ \\    
				\midrule    
				\multirow{5}{*}{$e_1$}     
				& $A_1$ & - & 0.0000 & 0.0000 & 0.0000 & 0.0000 & - & 0.0977 & 0.0121 & 0.0000 & 0.0000 & - & 0.0000 & 0.0000 & 0.0344 & 0.1029 \\    
				& $A_2$ & 0.0000 & - & 0.0000 & 0.1101 & 0.0000 & 0.0000 & - & 0.0000 & 0.0000 & 0.0000 & 0.0000 & - & 0.0000 & 0.0385 & 0.1006 \\    
				& $A_3$ & 0.1385 & 0.1817 & - & 0.2717 & 0.0627 & 0.0000 & 0.0000 & - & 0.0000 & 0.0000 & 0.0000 & 0.0000 & - & 0.0000 & 0.0675 \\    
				& $A_4$ & 0.0000 & 0.0000 & 0.0000 & - & 0.0000 & 0.0000 & 0.0167 & 0.0000 & - & 0.0000 & 0.0000 & 0.0000 & 0.0000 & - & 0.0853 \\    
				& $A_5$ & 0.1198 & 0.1592 & 0.0000 & 0.2517 & - & 0.0000 & 0.0422 & 0.0000 & 0.0324 & - & 0.0000 & 0.0000 & 0.0000 & 0.0000 & - \\    
				\midrule    
				\multirow{5}{*}{$e_2$}     
				& $A_1$ & - & 0.1272 & 0.2396 & 0.0000 & 0.0000 & - & 0.0000 & 0.0000 & 0.0000 & 0.0000 & - & 0.0886 & 0.1048 & 0.1519 & 0.1745 \\    
				& $A_2$ & 0.0000 & - & 0.1448 & 0.0000 & 0.0000 & 0.1311 & - & 0.0000 & 0.0473 & 0.0000 & 0.0000 & - & 0.0000 & 0.1035 & 0.1194 \\    
				& $A_3$ & 0.0000 & 0.0000 & - & 0.0000 & 0.0000 & 0.1214 & 0.0000 & - & 0.0000 & 0.0000 & 0.0000 & 0.0000 & - & 0.1281 & 0.0797 \\    
				& $A_4$ & 0.0000 & 0.1314 & 0.2071 & - & 0.0593 & 0.0000 & 0.0000 & 0.0000 & - & 0.0000 & 0.0000 & 0.0000 & 0.0000 & - & 0.0000 \\    
				& $A_5$ & 0.0000 & 0.0822 & 0.1967 & 0.0000 & - & 0.1218 & 0.0000 & 0.0000 & 0.0000 & - & 0.0000 & 0.0000 & 0.0000 & 0.0000 & - \\    
				\midrule    
				\multirow{5}{*}{$e_3$}     
				& $A_1$ & - & 0.0000 & 0.0000 & 0.0000 & 0.0000 & - & 0.0000 & 0.0000 & 0.0000 & 0.0000 & - & 0.2970 & 0.0000 & 0.0000 & 0.0000 \\    
				& $A_2$ & 0.1707 & - & 0.0000 & 0.1564 & 0.0000 & 0.0000 & - & 0.0000 & 0.0555 & 0.0000 & 0.0000 & - & 0.0000 & 0.0000 & 0.0000 \\    
				& $A_3$ & 0.2507 & 0.1119 & - & 0.2066 & 0.0000 & 0.1298 & 0.2054 & - & 0.2401 & 0.0000 & 0.0000 & 0.2950 & - & 0.0000 & 0.0000 \\    
				& $A_4$ & 0.1498 & 0.0000 & 0.0000 & - & 0.0000 & 0.0000 & 0.0000 & 0.0000 & - & 0.0000 & 0.0000 & 0.4121 & 0.1902 & - & 0.2160 \\    
				& $A_5$ & 0.2432 & 0.1118 & 0.0000 & 0.1954 & - & 0.1796 & 0.2676 & 0.0847 & 0.2772 & - & 0.0000 & 0.3160 & 0.0000 & 0.0000 & - \\    
				\bottomrule  
			\end{tabular}
		\end{minipage}
	}
\end{table}

\subsection{The second period}\label{Sec}
\setcounter{table}{0}
\begin{table}[H]
	\centering
	\caption{The probability distributions of the attribute evaluation results for alternative suppliers.(e=1)}
	\label{Tab:A2-1}
	\setlength{\tabcolsep}{4pt}
	\renewcommand{\arraystretch}{0.8}
	\begin{tabular}{ccccc|ccccc}
		\toprule
		\textbf{Alternative} & \textbf{Rating} & \textbf{$C_1$} & \textbf{$C_2$} & \textbf{$C_3$}
		& \textbf{Alternative} & \textbf{Rating} & \textbf{$C_1$} & \textbf{$C_2$} & \textbf{$C_3$} \\
		\midrule
		\multirow{5}{*}{$A_1$}
		& 1 & 0.291 & 0.198 & 0.133 & \multirow{5}{*}{$A_6$} & 1 & 0.004 & 0.097 & 0.259 \\
		& 2 & 0.126 & 0.135 & 0.193 & & 2 & 0.219 & 0.271 & 0.243 \\
		& 3 & 0.120 & 0.293 & 0.299 & & 3 & 0.320 & 0.051 & 0.081 \\
		& 4 & 0.184 & 0.248 & 0.250 & & 4 & 0.170 & 0.318 & 0.241 \\
		& 5 & 0.279 & 0.126 & 0.125 & & 5 & 0.287 & 0.263 & 0.176 \\
		\midrule
		\multirow{5}{*}{$A_2$}
		& 1 & 0.139 & 0.210 & 0.113 & \multirow{5}{*}{$A_7$} & 1 & 0.247 & 0.181 & 0.224 \\
		& 2 & 0.294 & 0.394 & 0.249 & & 2 & 0.229 & 0.430 & 0.084 \\
		& 3 & 0.430 & 0.021 & 0.263 & & 3 & 0.476 & 0.005 & 0.318 \\
		& 4 & 0.040 & 0.174 & 0.240 & & 4 & 0.005 & 0.196 & 0.262 \\
		& 5 & 0.097 & 0.201 & 0.135 & & 5 & 0.043 & 0.188 & 0.112 \\
		\midrule
		\multirow{5}{*}{$A_3$}
		& 1 & 0.046 & 0.234 & 0.083 & & & & & \\
		& 2 & 0.178 & 0.091 & 0.446 & & & & & \\
		& 3 & 0.353 & 0.300 & 0.097 & & & & & \\
		& 4 & 0.323 & 0.250 & 0.195 & & & & & \\
		& 5 & 0.100 & 0.125 & 0.179 & & & & & \\
		\bottomrule
	\end{tabular}
\end{table}

\begin{table}[H]
	\centering
	\caption{The probability distributions of the attribute evaluation results for alternative suppliers.(e=2)}
	\label{Tab:A2-2}
	\setlength{\tabcolsep}{4pt}
	\renewcommand{\arraystretch}{0.8}
	\begin{tabular}{ccccc|ccccc}
		\toprule
		\textbf{Alternative} & \textbf{Rating} & \textbf{$C_1$} & \textbf{$C_2$} & \textbf{$C_3$}
		& \textbf{Alternative} & \textbf{Rating} & \textbf{$C_1$} & \textbf{$C_2$} & \textbf{$C_3$} \\
		\midrule
		\multirow{5}{*}{$A_1$}
		& 1 & 0.130 & 0.413 & 0.004 & \multirow{5}{*}{$A_6$} & 1 & 0.036 & 0.196 & 0.219 \\
		& 2 & 0.279 & 0.004 & 0.303 & & 2 & 0.202 & 0.345 & 0.409 \\
		& 3 & 0.060 & 0.340 & 0.319 & & 3 & 0.430 & 0.107 & 0.004 \\
		& 4 & 0.360 & 0.123 & 0.343 & & 4 & 0.205 & 0.348 & 0.177 \\
		& 5 & 0.171 & 0.120 & 0.031 & & 5 & 0.127 & 0.004 & 0.191 \\
		\midrule
		\multirow{5}{*}{$A_2$}
		& 1 & 0.231 & 0.160 & 0.141 & \multirow{5}{*}{$A_7$} & 1 & 0.005 & 0.209 & 0.360 \\
		& 2 & 0.087 & 0.388 & 0.187 & & 2 & 0.276 & 0.441 & 0.022 \\
		& 3 & 0.516 & 0.188 & 0.366 & & 3 & 0.438 & 0.013 & 0.389 \\
		& 4 & 0.005 & 0.004 & 0.216 & & 4 & 0.206 & 0.018 & 0.209 \\
		& 5 & 0.161 & 0.260 & 0.090 & & 5 & 0.075 & 0.319 & 0.020 \\
		\midrule
		\multirow{5}{*}{$A_3$}
		& 1 & 0.262 & 0.171 & 0.030 & & & & & \\
		& 2 & 0.315 & 0.244 & 0.580 & & & & & \\
		& 3 & 0.181 & 0.279 & 0.035 & & & & & \\
		& 4 & 0.240 & 0.300 & 0.190 & & & & & \\
		& 5 & 0.002 & 0.006 & 0.165 & & & & & \\
		\bottomrule
	\end{tabular}
\end{table}

\begin{table}[H]
	\centering
	\caption{The probability distributions of the attribute evaluation results for alternative suppliers.(e=3)}
	\label{Tab:A2-3}
	\setlength{\tabcolsep}{4pt}
	\renewcommand{\arraystretch}{0.8}
	\begin{tabular}{ccccc|ccccc}
		\toprule
		\textbf{Alternative} & \textbf{Rating} & \textbf{$C_1$} & \textbf{$C_2$} & \textbf{$C_3$}
		& \textbf{Alternative} & \textbf{Rating} & \textbf{$C_1$} & \textbf{$C_2$} & \textbf{$C_3$} \\
		\midrule
		\multirow{5}{*}{$A_1$}
		& 1 & 0.323 & 0.319 & 0.145 & \multirow{5}{*}{$A_6$} & 1 & 0.004 & 0.005 & 0.297 \\
		& 2 & 0.223 & 0.003 & 0.295 & & 2 & 0.224 & 0.242 & 0.278 \\
		& 3 & 0.191 & 0.339 & 0.291 & & 3 & 0.405 & 0.088 & 0.053 \\
		& 4 & 0.053 & 0.132 & 0.264 & & 4 & 0.317 & 0.392 & 0.036 \\
		& 5 & 0.210 & 0.207 & 0.005 & & 5 & 0.050 & 0.273 & 0.336 \\
		\midrule
		\multirow{5}{*}{$A_2$}
		& 1 & 0.060 & 0.266 & 0.436 & \multirow{5}{*}{$A_7$} & 1 & 0.182 & 0.302 & 0.143 \\
		& 2 & 0.348 & 0.294 & 0.298 & & 2 & 0.217 & 0.376 & 0.016 \\
		& 3 & 0.286 & 0.128 & 0.176 & & 3 & 0.480 & 0.038 & 0.407 \\
		& 4 & 0.004 & 0.272 & 0.081 & & 4 & 0.054 & 0.201 & 0.313 \\
		& 5 & 0.302 & 0.040 & 0.009 & & 5 & 0.067 & 0.083 & 0.121 \\
		\midrule
		\multirow{5}{*}{$A_3$}
		& 1 & 0.003 & 0.124 & 0.206 & & & & & \\
		& 2 & 0.252 & 0.173 & 0.389 & & & & & \\
		& 3 & 0.266 & 0.208 & 0.026 & & & & & \\
		& 4 & 0.389 & 0.390 & 0.221 & & & & & \\
		& 5 & 0.090 & 0.105 & 0.158 & & & & & \\
		\bottomrule
	\end{tabular}
\end{table}

\begin{table}[H]
	\centering
	\rotatebox{90}{
		\begin{minipage}{22cm}
			\centering
			\footnotesize
			\caption{Stochastic Dominance Relation Matrix for Alternative Suppliers.(P=2)}
			\label{Tab:A2-4}
			\setlength{\tabcolsep}{1.5pt}
			\begin{tabular}{llccccccccccccccc}
				\toprule
				\multirow{2}{*}{\textbf{Expert}} & \multirow{2}{*}{}
				& \multicolumn{5}{c}{\textbf{Attribute $C_1$}}
				& \multicolumn{5}{c}{\textbf{Attribute $C_2$}}
				& \multicolumn{5}{c}{\textbf{Attribute $C_3$}} \\
				\cmidrule(lr){3-7} \cmidrule(lr){8-12} \cmidrule(lr){13-17}
				&
				& $A_1$ & $A_2$ & $A_3$ & $A_6$ & $A_7$
				& $A_1$ & $A_2$ & $A_3$ & $A_6$ & $A_7$
				& $A_1$ & $A_2$ & $A_3$ & $A_6$ & $A_7$ \\
				\midrule
				\multirow{5}{*}{$e_1$}
				& $A_1$ & - & - & - & - & - & - & 1.5SD & 1.25SD & - & - & - & - & - & 1.5SD & 1.25SD \\
				& $A_2$ & - & - & - & - & FSD & - & - & - & - & - & - & - & - & 1.5SD & 1.5SD \\
				& $A_3$ & 1.5SD & FSD & - & - & FSD & - & - & - & - & - & - & - & - & 1.5SD & - \\
				& $A_6$ & 1.25SD & FSD & FSD & - & FSD & 1.5SD & FSD & 1.5SD & - & FSD & - & - & - & - & - \\
				& $A_7$ & - & - & - & - & - & - & 1.75SD & - & - & - & - & - & - & 1.5SD & - \\
				\midrule
				\multirow{5}{*}{$e_2$}
				& $A_1$ & - & SSD & FSD & - & - & - & - & - & - & - & - & 1.5SD & 1.5SD & 1.5SD & FSD \\
				& $A_2$ & - & - & 1.5SD & - & - & 1.75SD & - & - & - & SSD & - & - & - & 1.5SD & FSD \\
				& $A_3$ & - & - & - & - & - & 1.5SD & - & - & 1.5SD & - & - & - & - & 1.25SD & 1.75SD \\
				& $A_6$ & SSD & 1.25SD & FSD & - & - & 1.75SD & - & - & - & - & - & - & - & - & - \\
				& $A_7$ & - & 1.25SD & FSD & - & - & - & - & - & - & - & - & - & - & - & - \\
				\midrule
				\multirow{5}{*}{$e_3$}
				& $A_1$ & - & - & - & - & - & - & - & - & - & - & - & 1.25SD & - & - & - \\
				& $A_2$ & FSD & - & - & - & 1.25SD & - & - & - & - & 1.25SD & - & - & - & - & - \\
				& $A_3$ & 1.25SD & 1.75SD & - & - & FSD & 1.5SD & FSD & - & - & FSD & - & FSD & - & - & - \\
				& $A_6$ & 1.25SD & SSD & - & - & 1.25SD & FSD & FSD & FSD & - & FSD & - & FSD & - & - & - \\
				& $A_7$ & SSD & - & - & - & - & - & - & - & - & - & FSD & FSD & 1.25SD & 1.5SD & - \\
				\bottomrule
			\end{tabular}
		\end{minipage}
	}
\end{table}

\begin{table}[H]
	\centering
	\rotatebox{90}{
		\begin{minipage}{22cm}
			\centering
			\footnotesize
			\caption{The Priority Degree Matrix for Alternative Suppliers. (P=2)}
			\label{Tab:A2-5}
			\setlength{\tabcolsep}{3pt}
			\begin{tabular}{llccccccccccccccc}    
				\toprule    
				
				\multirow{2}{*}{\textbf{Expert}} & \multirow{2}{*}{} 
				& \multicolumn{5}{c}{\textbf{Attribute $C_1$}} 
				& \multicolumn{5}{c}{\textbf{Attribute $C_2$}} 
				& \multicolumn{5}{c}{\textbf{Attribute $C_3$}} \\    
				\cmidrule(lr){3-7} \cmidrule(lr){8-12} \cmidrule(lr){13-17}    
				
				& 
				& $A_1$ & $A_2$ & $A_3$ & $A_6$ & $A_7$ 
				& $A_1$ & $A_2$ & $A_3$ & $A_6$ & $A_7$ 
				& $A_1$ & $A_2$ & $A_3$ & $A_6$ & $A_7$ \\    
				
				\midrule    
				
				\multirow{5}{*}{$e_1$} 
				& $A_1$ & - & 0.0000 & 0.0000 & 0.0000 & 0.0000 & - & 0.0977 & 0.0121 & 0.0000 & 0.0000 & - & 0.0000 & 0.0000 & 0.1106 & 0.0351 \\    
				& $A_2$ & 0.0000 & - & 0.0000 & 0.0000 & 0.1104 & 0.0000 & - & 0.0000 & 0.0000 & 0.0000 & 0.0000 & - & 0.0000 & 0.1080 & 0.0387 \\    
				& $A_3$ & 0.1385 & 0.1817 & - & 0.0000 & 0.2721 & 0.0000 & 0.0000 & - & 0.0000 & 0.0000 & 0.0000 & 0.0000 & - & 0.0706 & 0.0000 \\    
				& $A_6$ & 0.1681 & 0.2431 & 0.0751 & - & 0.3267 & 0.1327 & 0.1826 & 0.1455 & - & 0.1773 & 0.0000 & 0.0000 & 0.0000 & - & 0.0000 \\    
				& $A_7$ & 0.0000 & 0.0000 & 0.0000 & 0.0000 & - & 0.0000 & 0.0177 & 0.0000 & 0.0000 & - & 0.0000 & 0.0000 & 0.0000 & 0.0725 & - \\    
				
				\midrule    
				
				\multirow{5}{*}{$e_2$} 
				& $A_1$ & - & 0.1272 & 0.2396 & 0.0000 & 0.0000 & - & 0.0000 & 0.0000 & 0.0000 & 0.0000 & - & 0.0886 & 0.1048 & 0.1943 & 0.1897 \\    
				& $A_2$ & 0.0000 & - & 0.1448 & 0.0000 & 0.0000 & 0.1311 & - & 0.0000 & 0.0000 & 0.0409 & 0.0000 & - & 0.0000 & 0.1177 & 0.1435 \\    
				& $A_3$ & 0.0000 & 0.0000 & - & 0.0000 & 0.0000 & 0.1214 & 0.0000 & - & 0.0531 & 0.0000 & 0.0000 & 0.0000 & - & 0.0800 & 0.1666 \\    
				& $A_6$ & 0.0830 & 0.1458 & 0.2449 & - & 0.0000 & 0.0927 & 0.0000 & 0.0000 & - & 0.0000 & 0.0000 & 0.0000 & 0.0000 & - & 0.0000 \\    
				& $A_7$ & 0.0000 & 0.1178 & 0.2166 & 0.0000 & - & 0.0000 & 0.0000 & 0.0000 & 0.0000 & - & 0.0000 & 0.0000 & 0.0000 & 0.0000 & - \\    
				
				\midrule    
				
				\multirow{5}{*}{$e_3$} 
				& $A_1$ & - & 0.0000 & 0.0000 & 0.0000 & 0.0000 & - & 0.0000 & 0.0000 & 0.0000 & 0.0000 & - & 0.2970 & 0.0000 & 0.0000 & 0.0000 \\  
				& $A_2$ & 0.1707 & - & 0.0000 & 0.0000 & 0.1655 & 0.0000 & - & 0.0000 & 0.0000 & 0.0621 & 0.0000 & - & 0.0000 & 0.0000 & 0.0000 \\    
				& $A_3$ & 0.2507 & 0.1119 & - & 0.0000 & 0.2126 & 0.1298 & 0.2054 & - & 0.0000 & 0.2491 & 0.0000 & 0.2950 & - & 0.0000 & 0.0000 \\    
				& $A_6$ & 0.2237 & 0.1124 & 0.0000 & - & 0.1981 & 0.2119 & 0.3147 & 0.1375 & - & 0.3524 & 0.0000 & 0.3198 & 0.0000 & - & 0.0000 \\    
				& $A_7$ & 0.1326 & 0.0000 & 0.0000 & 0.0000 & - & 0.0000 & 0.0000 & 0.0000 & 0.0000 & - & 0.1734 & 0.4070 & 0.1836 & 0.2000 & - \\    
				
				\bottomrule  
			\end{tabular}
		\end{minipage}
	}
\end{table}

\subsection{The third period}\label{Thi}
\setcounter{table}{0}
\begin{table}[H]
	\centering
	\caption{The probability distributions of the attribute evaluation results for alternative suppliers.(e=1)}
	\label{Tab:A3-1}
	\setlength{\tabcolsep}{4pt}
	\renewcommand{\arraystretch}{0.8}
	\begin{tabular}{ccccc|ccccc}
		\toprule
		\textbf{Alternative} & \textbf{Rating} & \textbf{$C_1$} & \textbf{$C_2$} & \textbf{$C_3$}
		& \textbf{Alternative} & \textbf{Rating} & \textbf{$C_1$} & \textbf{$C_2$} & \textbf{$C_3$} \\
		\midrule
		\multirow{5}{*}{$A_1$}
		& 1 & 0.291 & 0.198 & 0.133 & \multirow{5}{*}{$A_5$} & 1 & 0.080 & 0.130 & 0.242 \\
		& 2 & 0.126 & 0.135 & 0.193 & & 2 & 0.196 & 0.415 & 0.299 \\
		& 3 & 0.120 & 0.293 & 0.299 & & 3 & 0.390 & 0.073 & 0.088 \\
		& 4 & 0.184 & 0.248 & 0.250 & & 4 & 0.133 & 0.199 & 0.198 \\
		& 5 & 0.279 & 0.126 & 0.125 & & 5 & 0.201 & 0.183 & 0.173 \\
		\midrule
		\multirow{5}{*}{$A_2$}
		& 1 & 0.139 & 0.210 & 0.113 & \multirow{5}{*}{$A_6$} & 1 & 0.004 & 0.097 & 0.259 \\
		& 2 & 0.294 & 0.394 & 0.249 & & 2 & 0.219 & 0.271 & 0.243 \\
		& 3 & 0.430 & 0.021 & 0.263 & & 3 & 0.320 & 0.051 & 0.081 \\
		& 4 & 0.040 & 0.174 & 0.240 & & 4 & 0.170 & 0.318 & 0.241 \\
		& 5 & 0.097 & 0.201 & 0.135 & & 5 & 0.287 & 0.263 & 0.176 \\
		\midrule
		\multirow{5}{*}{$A_3$}
		& 1 & 0.046 & 0.234 & 0.083 & \multirow{5}{*}{$A_7$} & 1 & 0.247 & 0.181 & 0.224 \\
		& 2 & 0.178 & 0.091 & 0.446 & & 2 & 0.229 & 0.430 & 0.084 \\
		& 3 & 0.353 & 0.300 & 0.097 & & 3 & 0.476 & 0.005 & 0.318 \\
		& 4 & 0.323 & 0.250 & 0.195 & & 4 & 0.005 & 0.196 & 0.262 \\
		& 5 & 0.100 & 0.125 & 0.179 & & 5 & 0.043 & 0.188 & 0.112 \\
		\bottomrule
	\end{tabular}
\end{table}

\begin{table}[H]
	\centering
	\caption{The probability distributions of the attribute evaluation results for alternative suppliers.(e=2)}
	\label{Tab:A3-2}
	\setlength{\tabcolsep}{4pt}
	\renewcommand{\arraystretch}{0.8}
	\begin{tabular}{ccccc|ccccc}
		\toprule
		\textbf{Alternative} & \textbf{Rating} & \textbf{$C_1$} & \textbf{$C_2$} & \textbf{$C_3$}
		& \textbf{Alternative} & \textbf{Rating} & \textbf{$C_1$} & \textbf{$C_2$} & \textbf{$C_3$} \\
		\midrule
		\multirow{5}{*}{$A_1$}
		& 1 & 0.130 & 0.413 & 0.004 & \multirow{5}{*}{$A_5$} & 1 & 0.095 & 0.129 & 0.209 \\
		& 2 & 0.279 & 0.004 & 0.303 & & 2 & 0.214 & 0.493 & 0.403 \\
		& 3 & 0.060 & 0.340 & 0.319 & & 3 & 0.404 & 0.006 & 0.005 \\
		& 4 & 0.360 & 0.123 & 0.343 & & 4 & 0.144 & 0.318 & 0.155 \\
		& 5 & 0.171 & 0.120 & 0.031 & & 5 & 0.143 & 0.054 & 0.228 \\
		\midrule
		\multirow{5}{*}{$A_2$}
		& 1 & 0.231 & 0.160 & 0.141 & \multirow{5}{*}{$A_6$} & 1 & 0.036 & 0.196 & 0.219 \\
		& 2 & 0.087 & 0.388 & 0.187 & & 2 & 0.202 & 0.345 & 0.409 \\
		& 3 & 0.516 & 0.188 & 0.366 & & 3 & 0.430 & 0.107 & 0.004 \\
		& 4 & 0.005 & 0.004 & 0.216 & & 4 & 0.205 & 0.348 & 0.177 \\
		& 5 & 0.161 & 0.260 & 0.090 & & 5 & 0.127 & 0.004 & 0.191 \\
		\midrule
		\multirow{5}{*}{$A_3$}
		& 1 & 0.262 & 0.171 & 0.030 & \multirow{5}{*}{$A_7$} & 1 & 0.005 & 0.209 & 0.360 \\
		& 2 & 0.315 & 0.244 & 0.580 & & 2 & 0.276 & 0.441 & 0.022 \\
		& 3 & 0.181 & 0.279 & 0.035 & & 3 & 0.438 & 0.013 & 0.389 \\
		& 4 & 0.240 & 0.300 & 0.190 & & 4 & 0.206 & 0.018 & 0.209 \\
		& 5 & 0.002 & 0.006 & 0.165 & & 5 & 0.075 & 0.319 & 0.020 \\
		\bottomrule
	\end{tabular}
\end{table}

\begin{table}[H]
	\centering
	\caption{The probability distributions of the attribute evaluation results for alternative suppliers.(e=3)}
	\label{Tab:A3-3}
	\setlength{\tabcolsep}{4pt}
	\renewcommand{\arraystretch}{0.8}
	\begin{tabular}{ccccc|ccccc}
		\toprule
		\textbf{Alternative} & \textbf{Rating} & \textbf{$C_1$} & \textbf{$C_2$} & \textbf{$C_3$}
		& \textbf{Alternative} & \textbf{Rating} & \textbf{$C_1$} & \textbf{$C_2$} & \textbf{$C_3$} \\
		\midrule
		\multirow{5}{*}{$A_1$}
		& 1 & 0.323 & 0.319 & 0.145 & \multirow{5}{*}{$A_5$} & 1 & 0.042 & 0.075 & 0.352 \\
		& 2 & 0.223 & 0.003 & 0.295 & & 2 & 0.127 & 0.258 & 0.215 \\
		& 3 & 0.191 & 0.339 & 0.291 & & 3 & 0.457 & 0.053 & 0.005 \\
		& 4 & 0.053 & 0.132 & 0.264 & & 4 & 0.310 & 0.416 & 0.111 \\
		& 5 & 0.210 & 0.207 & 0.005 & & 5 & 0.064 & 0.198 & 0.317 \\
		\midrule
		\multirow{5}{*}{$A_2$}
		& 1 & 0.060 & 0.266 & 0.436 & \multirow{5}{*}{$A_6$} & 1 & 0.004 & 0.005 & 0.297 \\
		& 2 & 0.348 & 0.294 & 0.298 & & 2 & 0.224 & 0.242 & 0.278 \\
		& 3 & 0.286 & 0.128 & 0.176 & & 3 & 0.405 & 0.088 & 0.053 \\
		& 4 & 0.004 & 0.272 & 0.081 & & 4 & 0.317 & 0.392 & 0.036 \\
		& 5 & 0.302 & 0.040 & 0.009 & & 5 & 0.050 & 0.273 & 0.336 \\
		\midrule
		\multirow{5}{*}{$A_3$}
		& 1 & 0.003 & 0.124 & 0.206 & \multirow{5}{*}{$A_7$} & 1 & 0.182 & 0.302 & 0.143 \\
		& 2 & 0.252 & 0.173 & 0.389 & & 2 & 0.217 & 0.376 & 0.016 \\
		& 3 & 0.266 & 0.208 & 0.026 & & 3 & 0.480 & 0.038 & 0.407 \\
		& 4 & 0.389 & 0.390 & 0.221 & & 4 & 0.054 & 0.201 & 0.313 \\
		& 5 & 0.090 & 0.105 & 0.158 & & 5 & 0.067 & 0.083 & 0.121 \\
		\bottomrule
	\end{tabular}
\end{table}

\begin{table}[H]
	\centering
	\rotatebox{90}{
		\begin{minipage}{20cm}
			\centering
			\footnotesize
			\caption{Stochastic Dominance Relation Matrix for Alternative Suppliers.(P=3)}
			\label{Tab:A3-4}
			\setlength{\tabcolsep}{1.5pt}
			\begin{tabular}{llcccccccccccccccccc}
				\toprule
				\multirow{2}{*}{\textbf{Expert}} & \multirow{2}{*}{}
				& \multicolumn{6}{c}{\textbf{Attribute $C_1$}}
				& \multicolumn{6}{c}{\textbf{Attribute $C_2$}}
				& \multicolumn{6}{c}{\textbf{Attribute $C_3$}} \\
				\cmidrule(lr){3-8} \cmidrule(lr){9-14} \cmidrule(lr){15-20}
				& & $A_1$ & $A_2$ & $A_3$ & $A_5$ & $A_6$ & $A_7$
				& $A_1$ & $A_2$ & $A_3$ & $A_5$ & $A_6$ & $A_7$
				& $A_1$ & $A_2$ & $A_3$ & $A_5$ & $A_6$ & $A_7$ \\
				\midrule
				\multirow{6}{*}{$e_1$}
				& $A_1$ & - & - & - & - & - & - & - & 1.5SD & 1.25SD & - & - & - & - & - & - & 1.25SD & 1.5SD & 1.25SD \\
				& $A_2$ & - & - & - & - & - & FSD & - & - & - & - & - & - & - & - & - & 1.25SD & 1.5SD & 1.5SD \\
				& $A_3$ & 1.5SD & FSD & - & 1.75SD & - & FSD & - & - & - & - & - & - & - & - & - & FSD & 1.5SD & - \\
				& $A_5$ & 1.75SD & FSD & - & - & - & FSD & - & 1.25SD & - & - & - & 1.25SD & - & - & - & - & - & - \\
				& $A_6$ & 1.25SD & FSD & FSD & FSD & - & FSD & 1.5SD & FSD & 1.5SD & FSD & - & FSD & - & - & - & - & - & - \\
				& $A_7$ & - & - & - & - & - & - & - & 1.75SD & - & - & - & - & - & - & - & 1.25SD & 1.5SD & - \\
				\midrule
				\multirow{6}{*}{$e_2$}
				& $A_1$ & - & SSD & FSD & - & - & - & - & - & - & - & - & - & - & 1.5SD & 1.5SD & 1.5SD & 1.5SD & FSD \\
				& $A_2$ & - & - & 1.5SD & - & - & - & 1.75SD & - & - & - & - & SSD & - & - & - & 1.75SD & 1.5SD & FSD \\
				& $A_3$ & - & - & - & - & - & - & 1.5SD & - & - & - & 1.5SD & - & - & - & - & 1.75SD & 1.25SD & 1.75SD \\
				& $A_5$ & - & 1.25SD & FSD & - & - & - & 1.75SD & - & - & - & - & - & - & - & - & - & FSD & - \\
				& $A_6$ & SSD & 1.25SD & FSD & 1.25SD & - & - & 1.75SD & - & - & - & - & - & - & - & - & - & - & - \\
				& $A_7$ & - & 1.25SD & FSD & 1.75SD & - & - & - & - & - & - & - & - & - & - & - & - & - & - \\
				\midrule
				\multirow{6}{*}{$e_3$}
				& $A_1$ & - & - & - & - & - & - & - & - & - & - & - & - & - & 1.25SD & - & - & - & - \\
				& $A_2$ & FSD & - & - & - & - & 1.25SD & - & - & - & - & - & 1.25SD & - & - & - & - & - & - \\
				& $A_3$ & 1.25SD & 1.75SD & - & - & - & FSD & 1.5SD & FSD & - & - & - & FSD & - & FSD & - & - & - & - \\
				& $A_5$ & 1.25SD & 1.75SD & - & - & - & 1.25SD & 1.25SD & FSD & 1.75SD & - & - & FSD & - & FSD & - & - & - & - \\
				& $A_6$ & 1.25SD & SSD & - & - & - & 1.25SD & FSD & FSD & FSD & FSD & - & FSD & - & FSD & - & - & - & - \\
				& $A_7$ & SSD & - & - & - & - & - & - & - & - & - & - & - & FSD & FSD & 1.25SD & 1.5SD & 1.5SD & - \\
				\bottomrule
			\end{tabular}
		\end{minipage}
	}
\end{table}

\begin{table}[H]
	\centering
	\rotatebox{90}{
		\begin{minipage}{22cm}
			\centering
			\footnotesize
			\caption{The Priority Degree Matrix for Alternative Suppliers. (P=3)}
			\label{Tab:A3-5}
			\setlength{\tabcolsep}{4pt}
			\begin{tabular}{llcccccccccccccccccc}
				\toprule
				\multirow{2}{*}{\textbf{Expert}} & \multirow{2}{*}{}
				& \multicolumn{6}{c}{\textbf{Attribute $C_1$}}
				& \multicolumn{6}{c}{\textbf{Attribute $C_2$}}
				& \multicolumn{6}{c}{\textbf{Attribute $C_3$}} \\
				\cmidrule(lr){3-8} \cmidrule(lr){9-14} \cmidrule(lr){15-20}
				& & $A_1$ & $A_2$ & $A_3$ & $A_5$ & $A_6$ & $A_7$
				& $A_1$ & $A_2$ & $A_3$ & $A_5$ & $A_6$ & $A_7$
				& $A_1$ & $A_2$ & $A_3$ & $A_5$ & $A_6$ & $A_7$ \\
				\midrule
				\multirow{6}{*}{$e_1$}
				& $A_1$ & - & 0.0000 & 0.0000 & 0.0000 & 0.0000 & 0.0000 & - & 0.0977 & 0.0121 & 0.0000 & 0.0000 & 0.0000 & - & 0.0000 & 0.0000 & 0.1124 & 0.1106 & 0.0351 \\
				& $A_2$ & 0.0000 & - & 0.0000 & 0.0000 & 0.0000 & 0.1104 & 0.0000 & - & 0.0000 & 0.0000 & 0.0000 & 0.0000 & 0.0000 & - & 0.0000 & 0.1102 & 0.1080 & 0.0387 \\
				& $A_3$ & 0.1385 & 0.1817 & - & 0.0553 & 0.0000 & 0.2721 & 0.0000 & 0.0000 & - & 0.0000 & 0.0000 & 0.0000 & 0.0000 & 0.0000 & - & 0.0612 & 0.0706 & 0.0000 \\
				& $A_5$ & 0.1267 & 0.1626 & 0.0000 & - & 0.0000 & 0.2551 & 0.0000 & 0.0544 & 0.0000 & - & 0.0000 & 0.0447 & 0.0000 & 0.0000 & 0.0000 & - & 0.0000 & 0.0000 \\
				& $A_6$ & 0.1681 & 0.2431 & 0.0751 & 0.0961 & - & 0.3267 & 0.1327 & 0.1826 & 0.1455 & 0.1447 & - & 0.1773 & 0.0000 & 0.0000 & 0.0000 & 0.0000 & - & 0.0000 \\
				& $A_7$ & 0.0000 & 0.0000 & 0.0000 & 0.0000 & 0.0000 & - & 0.0000 & 0.0177 & 0.0000 & 0.0000 & 0.0000 & - & 0.0000 & 0.0000 & 0.0000 & 0.0804 & 0.0725 & - \\
				\midrule
				\multirow{6}{*}{$e_2$}
				& $A_1$ & - & 0.1272 & 0.2396 & 0.0000 & 0.0000 & 0.0000 & - & 0.0000 & 0.0000 & 0.0000 & 0.0000 & 0.0000 & - & 0.0886 & 0.1048 & 0.1712 & 0.1943 & 0.1897 \\
				& $A_2$ & 0.0000 & - & 0.1448 & 0.0000 & 0.0000 & 0.0000 & 0.1311 & - & 0.0000 & 0.0000 & 0.0000 & 0.0409 & 0.0000 & - & 0.0000 & 0.1152 & 0.1177 & 0.1435 \\
				& $A_3$ & 0.0000 & 0.0000 & - & 0.0000 & 0.0000 & 0.0000 & 0.1214 & 0.0000 & - & 0.0000 & 0.0531 & 0.0000 & 0.0000 & 0.0000 & - & 0.0984 & 0.0800 & 0.1666 \\
				& $A_5$ & 0.0000 & 0.0916 & 0.2052 & - & 0.0000 & 0.0000 & 0.1179 & 0.0000 & 0.0000 & - & 0.0000 & 0.0000 & 0.0000 & 0.0000 & 0.0000 & - & 0.0280 & 0.0000 \\
				& $A_6$ & 0.0830 & 0.1458 & 0.2449 & 0.0599 & - & 0.0000 & 0.0927 & 0.0000 & 0.0000 & 0.0000 & - & 0.0000 & 0.0000 & 0.0000 & 0.0000 & 0.0000 & - & 0.0000 \\
				& $A_7$ & 0.0000 & 0.1178 & 0.2166 & 0.0592 & 0.0000 & - & 0.0000 & 0.0000 & 0.0000 & 0.0000 & 0.0000 & - & 0.0000 & 0.0000 & 0.0000 & 0.0000 & 0.0000 & - \\
				\midrule
				\multirow{6}{*}{$e_3$}
				& $A_1$ & - & 0.0000 & 0.0000 & 0.0000 & 0.0000 & 0.0000 & - & 0.0000 & 0.0000 & 0.0000 & 0.0000 & 0.0000 & - & 0.2970 & 0.0000 & 0.0000 & 0.0000 & 0.0000 \\
				& $A_2$ & 0.1707 & - & 0.0000 & 0.0000 & 0.0000 & 0.1655 & 0.0000 & - & 0.0000 & 0.0000 & 0.0000 & 0.0621 & 0.0000 & - & 0.0000 & 0.0000 & 0.0000 & 0.0000 \\
				& $A_3$ & 0.2507 & 0.1119 & - & 0.0000 & 0.0000 & 0.2126 & 0.1298 & 0.2054 & - & 0.0000 & 0.0000 & 0.2491 & 0.0000 & 0.2950 & - & 0.0000 & 0.0000 & 0.0000 \\
				& $A_5$ & 0.2333 & 0.0976 & 0.0000 & - & 0.0000 & 0.2079 & 0.1684 & 0.2579 & 0.0715 & - & 0.0000 & 0.2988 & 0.0000 & 0.3174 & 0.0000 & - & 0.0000 & 0.0000 \\
				& $A_6$ & 0.2237 & 0.1124 & 0.0000 & 0.0000 & - & 0.1981 & 0.2119 & 0.3147 & 0.1375 & 0.0765 & - & 0.3524 & 0.0000 & 0.3198 & 0.0000 & 0.0000 & - & 0.0000 \\
				& $A_7$ & 0.1326 & 0.0000 & 0.0000 & 0.0000 & 0.0000 & - & 0.0000 & 0.0000 & 0.0000 & 0.0000 & 0.0000 & - & 0.1734 & 0.4070 & 0.1836 & 0.2081 & 0.2000 & - \\
				\bottomrule
			\end{tabular}
		\end{minipage}
	}
\end{table}

\newcommand{\DOI}[1]{Doi: \href{https://doi.org/#1}{#1}}
\renewcommand{\citep}[1]{\textsuperscript{\citep*{#1}}}
\newcommand{\supcite}[1]{\textsuperscript{\citep*{#1}}}
\bibliographystyle{elsarticle-num-names}
\bibliography{fractional}

\end{document}